\documentclass[11pt,a4paper]{article}

\usepackage{amsmath}
\usepackage{amssymb}
\usepackage{hyperref}
\usepackage{graphicx}
\graphicspath{ {./figs/} }
\usepackage{bbold}
\usepackage{color}
\usepackage{geometry}
\usepackage{setspace}
\usepackage{xspace}
\usepackage{ulem}

\newcommand{\MeV}[0]{\text{MeV}}
\newcommand{\cm}[0]{\text{cm}}
\newcommand{\de}[0]{\text{d}}
\newcommand{\ket}[1]{\left| #1 \right>}
\newcommand{\bra}[1]{\left< #1 \right|}
\newcommand{\braket}[2]{\left< #1 | #2 \right>}
\newcommand{\peanuts}[0]{\textsf{PEANUTS}\xspace}
\newcommand{\Python}[0]{\textsf{Python}\xspace}
\newcommand{\yaml}[0]{\textsf{YAML}\xspace}

\usepackage{listings}
\usepackage{tikz}

\definecolor{solarized@base03}{HTML}{002B36}
\definecolor{solarized@base02}{HTML}{073642}
\definecolor{solarized@base01}{HTML}{586e75}
\definecolor{solarized@base00}{HTML}{657b83}
\definecolor{solarized@base0}{HTML}{839496}
\definecolor{solarized@base1}{HTML}{93a1a1}
\definecolor{solarized@base2}{HTML}{EEE8D5}
\definecolor{solarized@base3}{HTML}{FDF6E3}
\definecolor{solarized@yellow}{HTML}{B58900}
\definecolor{solarized@orange}{HTML}{CB4B16}
\definecolor{solarized@red}{HTML}{DC322F}
\definecolor{solarized@magenta}{HTML}{D33682}
\definecolor{solarized@violet}{HTML}{6C71C4}
\definecolor{solarized@blue}{HTML}{157AC1}
\definecolor{solarized@cyan}{HTML}{2AA198}
\definecolor{solarized@green}{HTML}{859900}
\definecolor{darkred}{HTML}{550003}
\definecolor{darkgreen}{HTML}{00AA00}
\definecolor{orchid}{HTML}{AF06F5}

\lstdefinestyle{python}
{
  language=Python,
  basicstyle=\footnotesize\ttfamily,
  basewidth={0.53em,0.44em},
  numbers=none,
  tabsize=2,
  breaklines=true,
  escapeinside={@}{@},
  showstringspaces=false,
  numberstyle=\tiny\color{solarized@base01},
  keywordstyle=\color{solarized@blue},
  stringstyle=\color{solarized@red}\ttfamily,
  identifierstyle=\color{solarized@blue},
  commentstyle=\color{purple},
  emphstyle=\color{green},
  frame=single,
  rulecolor=\color{solarized@base2},
  rulesepcolor=\color{solarized@base2},
  literate = {~}{\customtilde}1
             {\ as\ }{{\color{blue}\ as\ \color{black}}}3
             {.set}{{\color{black}.}{\color{darkred}set}}4
}
\lstnewenvironment{lstpy}{\lstset{style=python}}{\lstset{style=python}}
\newcommand\py[1]{{\lstset{style=python}\lstinline!#1!\lstset{style=python}}}

\begin{document}

\noindent \today
\hfill TTK-23-001, TTP23-012

\vskip 0.4cm

\begin{center}
\bigskip
{\huge\bf
\begin{spacing}{1.1}
PEANUTS :\\
a software for the automatic computation of solar neutrino flux and its propagation within Earth
\end{spacing}
}

\end{center}

\vskip 0.4cm

\renewcommand*{\thefootnote}{\fnsymbol{footnote}}
{\large
\noindent Tom\'as E. Gonzalo$^{\dagger,1}$ and Michele Lucente$^{\star,2,3}$
\\[3mm]
{\it{
$^1$ Institute for Theoretical Particle Physics (TTP), Karlsruhe Institute of Technology (KIT), 76128 Karlsruhe, Germany \\
$^2$ Institute for Theoretical Particle Physics and Cosmology (TTK), RWTH Aachen University, D-52056 Aachen, Germany \\
$^3$ Dipartimento di Fisica e Astronomia, Universit\`a di Bologna, via Irnerio 46, 40126 Bologna, Italy}
}}
\\[3mm]
{$^{\dagger}$tomas.gonzalo@kit.edu, $^{\star}$michele.lucente@unibo.it}
\vskip 0.4cm

\renewcommand*{\thefootnote}{\arabic{footnote}}
\setcounter{footnote}{0}

\begin{abstract}
We present \peanuts (Propagation and Evolution of Active NeUTrinoS), an open-source Python package for the automatic computation of solar neutrino spectra and active neutrino propagation through Earth. \peanuts is designed to be \emph{fast}, by employing analytic formulae for the neutrino propagation through varying matter density, and \emph{flexible}, by allowing the user to input arbitrary solar models, custom Earth density profiles and general detector locations. It provides functionalities for a fully automated simulation of solar neutrino fluxes at a detector, as well as access to individual routines to perform more specialised computations. The software has been extensively tested against the results of the SNO experiment, providing excellent agreement with their results. In addition, the present text contains a pedagogical derivation of the relations needed to compute the oscillated solar neutrino spectra, neutrino propagation through Earth and nadir exposure of an experiment.
\end{abstract}

\thispagestyle{empty}

\pagebreak

\begin{small}
\tableofcontents
\end{small}

\newpage


\section{Introduction}
Solar neutrinos represent the most abundant source of neutrinos on Earth, with a flux of the order of $~6 \times 10^{10}\ \text{cm}^{-2}\ s^{-1}$~\cite{Bahcall:1989ks}. Even though only a small fraction of this flux is actually detectable, due to many production channels resulting in neutrinos with energies below the typical experimental detection thresholds~\cite{Bahcall:2004pz}, solar neutrinos provide an invaluable source of information for the study of neutrino properties, solar dynamics and Earth internal structure. Historically, solar neutrinos provided the first hints for the non-conservation of leptonic flavours in neutrino propagation~\cite{Bahcall:1981br}, when in 1968 the Homestake experiment reported far less solar electron neutrino events~\cite{Davis:1968cp} than the number expected from the recently developed solar models~\cite{Bahcall:1963ohf, Sears:1964zz, Pochoda:1964ana, Bahcall:1964gx, ez06000k, ez07000g, 1967ApJ...150..723B, 1967ApJ...150..725S, Bahcall:1968jvj}. This so-called \emph{solar neutrino problem} was only solved in 2001, when the SNO collaboration released~\cite{SNO:2001kpb} the measurement of the total flux of active ${}^8$B neutrinos, that resulted in a close agreement with solar models predictions and implied that the origin of the discrepancy had to be tracked to new physics effects in the neutrino sector. It is now firmly established that the deficit is due to neutrino oscillations~\cite{Maki:1962mu, Pontecorvo:1967fh, Gribov:1968kq}, with  oscillation parameters as inferred by solar neutrino experiments well in agreement with data from other neutrino sources~\cite{Esteban:2020cvm, deSalas:2020pgw, Capozzi:2017ipn}. 

Nowadays, solar neutrinos continue providing invaluable information, being the only known probes that can directly test the interior structure and dynamics of the Sun, thereby providing strong constraints on the solar models parameters~\cite{BOREXINO:2020aww, Borexino:2017rsf, BOREXINO:2018ohr}, or probe the internal solar dynamics on long time-scales~\cite{Yano:2020aap}. More exotic production mechanisms such as  neutrino emission during intense solar flares~\cite{IceCube:2021jwt, Super-Kamiokande:2022yrk}, neutrino production from cosmic-rays scattering in the solar atmosphere~\cite{IceCube:2021koo} and neutrinos from dark matter annihilation in the Sun~\cite{In:2017kcf} can also be probed. On the other hand, the theoretically established solar models allow to study and constrain neutrino properties, from standard oscillation parameters~\cite{Super-Kamiokande:2002ujc, Bahcall:2003ce} to more hypothetical scenarios such as non-standard interactions~\cite{Super-Kamiokande:2022lyl} or finite magnetic moment~\cite{Super-Kamiokande:2020frs}.

In all these scenarios, an important wealth of statistical information is encoded in the energy spectra for the individual neutrino flavours. Due to neutrino oscillations, however, the spectra on Earth generally differ from the ones at neutrino production, and a proper account of the oscillation dynamics is mandatory in order to compare a theoretical model with data. The oscillation dynamics is in general non-trivial, including three different regimes and corresponding matter effects~\cite{Wolfenstein:1977ue, Mikheyev:1985zog}: propagation within the slowly-changing\footnote{\emph{Slow} or \emph{fast} here refers to the importance of variation of matter density over a neutrino oscillation length scale.} matter density within the Sun, propagation in vacuum between Sun surface and Earth, and propagation within Earth featuring a fastly-evolving matter density profile.

In this work we present \peanuts (Propagation and Evolution of Active NeUTrinoS), an open-source, fast and flexible package to compute the neutrino oscillation dynamics in all the above-mentioned regimes. The emphasis in developing the software has been put on both \emph{performance} and \emph{flexibility}: \peanuts computes the coherent neutrino propagation inside Earth analytically, completely removing the need for  time-consuming numerical integrations. Moreover, the user can input any arbitrary solar model, as well as any custom Earth matter density profile, and simulate experiments at any latitude and underground depth. The software can perform the full chain of computations to simulate the expected neutrino spectra for a given solar model, Earth matter density and detector location, or its modules can be called individually, to compute for instance the evolved neutrino state after Earth crossing given an arbitrary initial (coherent or incoherent) state, or the solar angle distribution for an experiment taking data between two arbitrary days of the year.
\peanuts\footnote{https://github.com/michelelucente/PEANUTS} is provided as an open-source \Python package, under the GPL-3.0 license. It can be run standalone in various different ways, as well as interfaced from other frameworks. Details of the various software requirements and dependencies, as well as instructions on how to use \peanuts can be found in Section \ref{sec:quickstart}, whereas the specific functions and classes that perfom the computations will be scattered throught whenever the physical quantity computed is introduced. We have extensively validated our implementation against the results of the SNO experiment~\cite{FiuzadeBarros:2011qna,SNO:2006odc}, details of which will also be located where useful, including a final comparison of likelihood contours for our simulation of the experiment with the probabilities computed by \peanuts.

This document is thus structured as follows. Section \ref{sec:solar} describes the theoretical background for the computation of the solar neutrino flux and the propagation of the neutrinos from the Sun. In Section \ref{sec:earth} we describe the effect of Earth regeneration of the neutrino flux, i.e. the propagation of neutrinos through the Earth, followed by a detailed explanation of our fast approximation for the neutrino propagation Hamiltonian in Section \ref{sec:propag}. Section \ref{sec:time_integration} details the time integration over the exposure of a given experiment. For interested users, Section \ref{sec:quickstart} provides a quick start guide to \peanuts and a explanations on how to reproduce our validation procedure. Lastly, we provide our conclusions and outlook in Section \ref{sec:conclusions}.

\section{Solar neutrino flux}
\label{sec:solar}

\subsection{Neutrino survival probability at Sun surface}
\begin{figure}[t]
\includegraphics[width=0.5\textwidth]{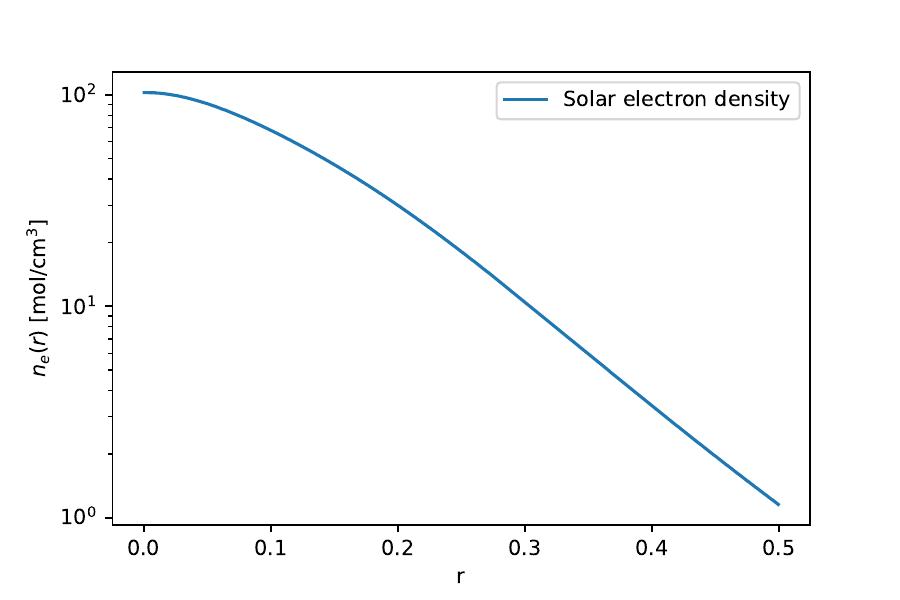}
\includegraphics[width=0.5\textwidth]{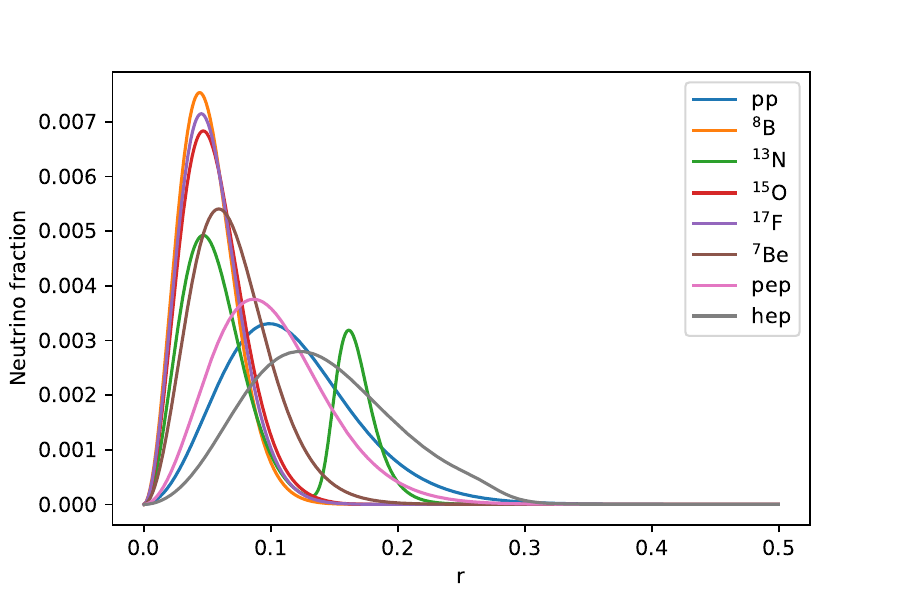}
\caption{Solar density (left) and fraction of solar neutrinos produced (right), as a function of relative radius $r \equiv R/R_\odot$, from Solar model BS05(AGS,OP)~\cite{Bahcall:2004pz}.}
\label{fig:BS05}
\end{figure}

Solar neutrinos are produced over a wide region within the Sun, making the incertitude over the production point much larger than the typical detectors size (in fact, much larger than the Earth itself. See e.g. Fig.~\ref{fig:BS05})~\cite{Bahcall:2000nu}. This feature implies that the solar neutrino flux at Earth is given by an \emph{incoherent} superposition of neutrino mass eigenstates~\cite{Mikheev:1987wa, Lisi:1997yc}, whose composition remains constant as the flux propagates in vacuum\footnote{The oscillation dynamics would be different in case of a coherent flux~\cite{Bruss:1988fr}.}. If the neutrino oscillation length-scale is much smaller than the scale over which matter density varies significantly, neutrino oscillations in matter proceed adiabatically~\cite{Mikheev:1986if}; the values of the neutrino oscillation parameters inferred by global fits of neutrino data~\cite{Esteban:2020cvm, deSalas:2020pgw, Capozzi:2017ipn} imply that the adiabatic regime is realised for solar neutrinos as they propagate from the production point towards the Sun surface. In the adiabatic approximation, the neutrino flavour composition at the Sun surface only depends on the matter effects at neutrino production point. 

Given $U=U(\theta_{12}, \theta_{13}, \theta_{23}, \delta)$ the PMNS mixing matrix in vacuum (cf. eq.s~(\ref{eq:pmns}, \ref{eq:PMNS_matrices})), it is possible to define an analogous matrix  $T$ that diagonalises the neutrino Hamiltonian in matter by simply replacing the vacuum values of $\theta_{12}, \theta_{13}$ with their matter-rotated ones~\cite{Denton:2016wmg, Denton:2018hal}\footnote{See also \cite{Ioannisian:2018qwl, Denton:2018mop}.}:
\begin{eqnarray}
\cos 2 \widetilde{\theta}_{13} & = & \frac{ (\cos 2\theta_{13} -A/\Delta m^2_{ee}) } 
{  \sqrt{(\cos 2\theta_{13}-A/\Delta m^2_{ee})^2 +  \sin^22\theta_{13}  ~ }}, 
 \label{eq:th13_matter}  
 \end{eqnarray}
 \begin{eqnarray}
 \cos 2 \widetilde{\theta}_{12} & = &  \frac{ ( \cos 2\theta_{12} 
 - A^{\,\prime}  /\Delta m^2_{21} ) } {  \sqrt{(\cos 2\theta_{12} 
 -A^{\,\prime} /\Delta m^2_{21})^2 ~+~
  \sin^2 2 \theta_{12} \cos^2( \widetilde{\theta}_{13}-\theta_{13})~~}  }, \label{eq:th12_matter} 
  \end{eqnarray}

	
%
	
where $\theta_{12}, \theta_{13}$ are the PMNS mixing angles in vacuum, $\widetilde{\theta}_{12},\widetilde{\theta}_{13}$ the corresponding ones in matter,
\begin{equation}
	A = 2 E V = 2 \sqrt{2} E G_F n_e
\end{equation}
represents the matter potential for a neutrino with energy $E$ travelling in a medium with electron density $n_e$, 
$G_F$ is the Fermi constant,
\begin{eqnarray}
    A^{\,\prime}    \equiv   A \, \cos^2 \widetilde{\theta}_{13} +\Delta m^2_{ee} \sin^2  ( \widetilde{\theta}_{13}-\theta_{13} )
\end{eqnarray}
is the $\theta_{13}$-modified matter potential and 
\begin{eqnarray}
    \Delta m^2_{ee} \equiv \cos^2 \theta_{12} \Delta m^2_{31} + \sin^2 \theta_{12} \Delta m^2_{32}.
\end{eqnarray}

The matrix $T$ is then simply defined as
\begin{eqnarray}\label{Tmatrix}
    T\left(E, n_e\right) = U(\widetilde{\theta}_{12}, \widetilde{\theta}_{13}, \theta_{23}, \delta).
\end{eqnarray}
Notice that the matrix $T$ depends both on local matter density and energy of the produced neutrino.

Numerically (cf. eq.s 4.17, 4.18 in \cite{Fantini:2018itu})
\begin{eqnarray}
	V &=& \sqrt{2} G_F n_e = \frac{3.868 \times 10^{-7}}{\text{m}} \times \frac{n_e}{\text{mol}/\text{cm}^3},\\
	k &=& \frac{\Delta m^2}{2 E} = \frac{2.533}{\text{m}} \times \frac{\Delta m^2}{\text{eV}^2} \times \frac{\MeV}{E},
\end{eqnarray}
where the wavenumber $k$ is useful in the relation
\begin{equation}
	\frac{A}{\Delta m^2} = \frac{2 \sqrt{2} E G_F n_e}{\Delta m^2} = \frac{V}{k}.
\end{equation}

The probability of producing a neutrino mass eigenstate $i$ in matter is thus $\left| T_{\alpha i}\left( E, n_e \right)\right|^2$, where $\alpha$ is the flavour of the charged lepton entering the neutrino production vertex.

\peanuts assumes the validity of the adiabatic regime when computing the solar neutrino flux, as this guarantees a fast computation and is an excellent approximation for the neutrino oscillation parameters realised in the standard 3-flavour mixing scheme~\cite{Kuo:1989qe,Bethe:1991zq,Harley:1992an,Akhmedov:2004rq,FiuzadeBarros:2011qna}. Hence, it provides a function for the computation of  the matrix $T$ in eq.\eqref{Tmatrix}, 
with signature

\begin{lstpy}
 Tei(th12,th13,DeltamSq21,DeltamSq3l,E,ne)
\end{lstpy}
where the arguments correspond, respectively,  to $\theta_{12}$, $\theta_{13}$, $\Delta m_{21}^2$ (eV$^2$), $\Delta m_{3\ell}^2$ (eV$^2$), $E$ (MeV) and $n_e$ (mol/cm$^3$). Note that in \peanuts the input variable $\Delta m_{3\ell}$ has different meanings according to the ordering of neutrino mass eigenstates. In normal ordering (NO) $l=1$, and thus $\Delta m_{31}^2 = \Delta m_{3\ell}^2$ and $\Delta m_{32}^2 = \Delta m_{3\ell}^2 - \Delta m_{21}^2$, whereas for inverted ordering (IO) $l=2$, so $\Delta m_{32}^2 = \Delta m_{3\ell}^2$ and $\Delta m_{31}^2 = \Delta m_{3\ell}^2 + \Delta m_{21}^2$. \peanuts also provides accessible functions for the mixing angles in matter $\widetilde{\theta}_{12}$ and $\widetilde{\theta}_{13}$, as well as useful quantities such as $\Delta m_{ee}^2$ and the ratio $V/k$, in the form of the functions
\begin{lstpy}
th12_M(th12,th13,DeltamSq21,DeltamSq3l,E,ne)
th13_M(th12,th13,DeltamSq21,DeltamSq31,E,ne)
DeltamSqee(th12,DeltamSq21,DeltamSq3l)
Vk(Deltam2,E,ne)
\end{lstpy}

In the adiabatic approximation neutrinos evolve as pure mass eigenstates within the Sun. For a fixed value of neutrino energy $E$, the flux composition  at Sun surface is given by the average over the neutrino production points inside the Sun. Assuming spherical symmetry, if $f(r)$ is the fraction of neutrinos produced at point $r \equiv R/R_\odot$, where $R_\odot$ is the solar radius and $R$ the distance from the center of the Sun, the probability of a solar neutrino with energy $E$ to emerge as mass eigenstate $i$ is
\begin{equation}
	P_{\nu_e \rightarrow \nu_i}^\odot(E) = \int_0^1 \de r \left| T_{e i}\left( E, n_e(r) \right)\right|^2 f(r),
\label{solarfluxmass}
\end{equation}
with the normalization
\begin{equation}
	\int_0^1 \de r f(r) = 1.
\end{equation}

The computation of $f(r)$ assumes a specific solar model. In our validation of \peanuts  we use the BS05(AGS,OP) model~\cite{Bahcall:2004pz}\footnote{Numerical data are available at \url{http://www.sns.ias.edu/~jnb/}}, which is one of the models assumed by the SNO collaboration in their neutrino oscillation fit~\cite{FiuzadeBarros:2011qna}.  Fig.~\ref{fig:BS05}-left shows the value of solar matter density $n_e(r)$ as a function of solar radius $r$ in the BS05(AGS,OP) model, while Fig.~\ref{fig:BS05}-right shows the neutrino production fractions for different production chains. Note that the electron density in~\cite{Bahcall:2004pz} is given in units of $\frac{1}{\cm^3 N_A} = \frac{\text{mol}}{\cm^3\ 6.022 \cdot 10^{23} }.$

The choice of a particular solar model is encoded in \peanuts as the \Python class \py{SolarModel}. Through the constructor of the class, of signature,
\begin{lstpy}
 SolarModel(solar_model_file=None, flux_file=None, spectrum_files=None, fluxrows=None, fluxcols=None, fluxscale=None, distrow=None, radiuscol=None, densitycol=None, fractioncols=None)
\end{lstpy}
one can select the location of a file describing the chosen solar model with the optional argument \py{solar_model_file}. By default, when no other file location is provided, \peanuts will assume the use of the BS16(AGSS09) model~\cite{Vinyoles:2016djt}. At the time of publication, \peanuts can also work out of the box with the B16 (GS98)~\cite{Vinyoles:2016djt} model, the BS05(AGS,OP) and BS05(OP) models~\cite{Bahcall:2004pz}, and the BP00~\cite{Bahcall:2000nu} solar model. The user is nevertheless encouraged to implement their own solar model (and thus neutrino fraction distribution $f(r)$), by providing a custom solar model file via the \py{solar_model_file} optional argument and/or flux file with \py{flux_file} (they will be assumed to be the same if the latter argument is missing), but in such case one must also specify the rows and columns where the relevant information in the files can be located. The options \py{fluxrows} and \py{fluxcols} sets the rows and columns in \py{flux_file} where one can find the total neutrino flux per fraction, which can be either dictionaries or real numbers, but at least one of the two must be a dictionary with the names of the fractions of interest and the corresponding row or column. If the fluxes must be rescaled, the option \py{fluxscale} can be provided, which can be a real number, for blanket rescaling of all fluxes, or a dictionary with different rescaling for each fraction. The option \py{distrow} points to the first row of the table in \py{solar_model_file} containing the distributions, i.e. the radius, density and fraction samples, whose columns must also be specified with the options \py{radiuscol}, \py{densitycol} and \py{fractioncols}, respectively, with the latter a dictionary of names and columns of the fractions of interest. Naturally all of these optional arguments must be provided if the solar model file is not known to \peanuts.

In addition to allowing the selection of different solar models, \peanuts also allows the use of different energy spectra for the various neutrino fluxes. By default, the following are provided and taken from: $pp$ and $hep$~\cite{Bahcall:1997eg}, $^8B$~\cite{Ortiz:2000nf}, ${}^{13}N$, ${}^{15}O$ and ${}^{17}F$~\cite{Bahcall:1987jc}, $^7Be$~\cite{Bahcall:1994cf}. Different spectra can be provided via the optional argument \py{spectrum_files} in the constructor for the \py{SolarModel} class. It should be noted that spectra are assumed in \peanuts to be normalised to 1, so the user should make sure to renormalise the spectrum accordingly, as it is the case, for instance for the $^8B$ spectrum from \cite{Winter:2004kf}, which is normalized to 1000.

In addition to the neutrino fraction distributions and fluxes, the \py{SolarModel} class in \peanuts provides the density of the Sun at various radius samples, necessary for computing the probability of neutrinos at Sun exit from eq.~\eqref{solarfluxmass}. This probability, or equivalenty the weight of neutrino mass eigenstates in the surface of the Sun, is computed by \peanuts with the function
\begin{lstpy}
 solar_flux_mass(th12, th13, DeltamSq21, DeltamSq3l, E, radius_samples, density, fraction)
\end{lstpy}

\subsection{Neutrino propagation from the Sun}

Being it an incoherent flux, the fraction of mass eigenstates within the solar neutrino flux remains constant as long as neutrinos propagate in the vacuum, on their path from Sun surface to Earth. Here we adopt the convention that the unitary PMNS mixing matrix $U$ defines the change of basis between mass and flavour neutrino \emph{fields}
\begin{equation}\label{eq:nu_field_mixing}
	\nu_\alpha (x) = U_{\alpha i}\ \nu_i (x),
\end{equation}
with $\alpha = e,\mu,\tau$ the flavour indexes, $i=1,2,3$ the mass ones. With this convention, a neutrino \emph{state} produced at the origin results from the linear superposition of mass states\footnote{See \cite{Fantini:2018itu} for a comprehensive discussion on the relations between fields, plane waves and momentum eigenstates.}
\begin{equation}\label{eq:nu_state_mixing}
	\ket{\nu_\alpha, x=0} = U_{\alpha i}^* \ket{\nu_i, x=0},
\end{equation}
implying that the probability of observing a neutrino of flavour $\alpha$ from a mass eigenstate $i$ is given by $|U_{\alpha i}^*|^2$. Thus, the probability for a solar neutrino to manifest as flavour $\alpha$ is given by
\begin{equation}
	P^S_{\nu_e \rightarrow \nu_\alpha}(E) = |U_{\alpha i}^*|^2 P^\odot_{\nu_e \rightarrow \nu_i}(E),
\label{Psolar}
\end{equation}
and we assume throughout this paper that repeated indexes are summed.

In \peanuts the PMNS matrix is implemented as the \py{PMNS} class, which is constructed from the mixing angles $\theta_{ij}$ and CP phase $\delta_{\textrm{CP}}$ as
\begin{lstpy}
 PMNS(th12,th13,th23,d)
\end{lstpy}

This \py{PMNS} provides accessor functions for all mixings parameters, as well as other useful quantities such as the orthogonal/unitary matrices $R_{12}$, $R_{13}$, $R_{23}$ and $\Delta$, which allow the PMNS mixing matrix $U$ to be expressed as
\begin{equation}\label{eq:pmns}
	U = R_{23} \Delta R_{13} \Delta^* R_{12},
\end{equation}
and are defined as

\begin{equation}\label{eq:PMNS_matrices}
	\begin{array}{ll}
		R_{23} = \left( \begin{array}{ccc}
		1 & 0 & 0 \\
		0 & c_{23} & s_{23} \\
		0 & -s_{23} & c_{23}
	\end{array} \right), &
	R_{13} = \left( \begin{array}{ccc}
		c_{13} & 0 & s_{13}\\
		0 & 1 & 0 \\
		-s_{13} & 0 & c_{13}
	\end{array} \right), \\
			R_{12} = \left( \begin{array}{ccc}
		c_{12} & s_{12} & 0 \\
		-s_{12} & c_{12} & 0 \\
		0 & 0 & 1
	\end{array} \right), &
	\Delta = \left( \begin{array}{ccc}
		1 & 0 & 0\\
		0 & 1 & 0 \\
		0 & 0 & e^{i\delta}
	\end{array} \right).
	\end{array}
\end{equation}

This decomposition of the PMNS matrix will be useful further down when computing the propagation Hamiltonian through Earth. Finally, in order to compute the solar probability for flavour eigenstates with \peanuts, one can use the function
\begin{lstpy}
 PSolar(pmns,DeltamSq21,DeltamSq3l,E,radius_samples,density,fraction)
\end{lstpy}
which simply implements eq. \eqref{Psolar} using a \py{PMNS} object and calling the  \py{solar_flux_mass} function, and returns a list of the probability for each flavour eigenstate.

\section{Neutrino propagation through Earth}
\label{sec:earth}

\subsection{Probability of transition through matter}

If the neutrino flux from the Sun crosses the Earth (or any finite density matter in general) the probabilities are modified, since the propagation eigenstates in matter differ from the vacuum ones.

In general, a generic neutrino state at time\footnote{Since we always assume ultrarelativistic neutrinos, we can interchangeably use traveled distance $x$ or elapsed time $t$ to identify the evolved neutrino state.} $t$ can be expressed in terms of the state at time $t_0$ by evolving it with an appropriate evolutor operator $\hat{\mathcal{U}}(t,t_0)$
\begin{equation}\label{eq:evolutor_op}
    \ket{\nu, t} = \hat{\mathcal{U}}(t,t_0)\ket{\nu, t_0}. 
\end{equation}
The generic state $\ket{\nu,t}$ can be expressed as a linear superposition of pure flavour eigenstates, 
\begin{equation}
    \ket{\nu, t} = c_\alpha (t) \ket{\nu_\alpha},
    \label{eq:general_flavour_state}
\end{equation}
where $c_\alpha (t)$ are complex numbers, implying that the probability of it to interact as a neutrino with flavour $\alpha$ at time $t$ will be given by $|c_\alpha(t)|^2$.
From Eq.~\ref{eq:evolutor_op} it follows that the evolved probability amplitudes are given by
\begin{equation}
    c_\alpha (t) = \braket{\nu_\alpha}{\nu,t} = \bra{\nu_\alpha} \hat{\mathcal{U}}(t,t_0) \ket{\nu_\beta} \braket{\nu_\beta}{\nu,t_0}  = \mathcal{U}_{\alpha \beta}(t,t_0) c_\beta (t_0),
    \label{eq:evolved_flavour_coeff}
\end{equation}
where $\mathcal{U}_{\alpha \beta}(t,t_0)$ are the matrix elements of the evolutor operator in flavour basis.
 The determination of the evolutor operator $\mathcal{U}$ is in general a non-trivial problem, and will be discussed in detail in the following sections; for the moment let us assume we know a closed form expression for it.

A mass eigenstate expressed as linear combination of flavour eigenstates is
\begin{equation}
	\ket{\nu_i} = U_{i \alpha}^T \ket{\nu_\alpha} = U_{\alpha i} \ket{\nu_\alpha},
\end{equation}
which implies a transition amplitude from (evolved) mass to flavour eigenstate
\begin{equation}
	\braket{\nu_\alpha}{\nu_i, t} = \mathcal{U}_{\alpha \beta} (t, t_0) U_{\beta i}.
\end{equation}

Putting everything together, the final probability for a solar neutrino to manifest as $\alpha$ flavour while crossing the Earth is given by
\begin{equation}\label{eq:sun_earth_probability}
	P_\alpha^{SE}(t, E) = \left| \mathcal{U}_{\alpha \beta}(t, t_0) U_{\beta i} \right|^2  P_{\nu_e \rightarrow \nu_i}^\odot(E),
\end{equation}
where $t_0$ is defined at the time of neutrino crossing the Earth surface.
The interpretation of Eq.~(\ref{eq:sun_earth_probability}) is the following: $U_{\beta i}$ are the coefficients of the mass eigenstate $i$ expressed as linear combination of flavour eigenstates, $\ket{\nu_i} = U_{\beta i} \ket{\nu_\beta}$, and $\mathcal{U}_{\alpha \beta}(t, t_0) U_{\beta i} = \braket{\nu_\alpha}{\nu_i, t}$ is the transition amplitude from the evolved mass eigenstate $i$ to interaction eigenstate $\alpha$. Finally, each probability $|\braket{\nu_\alpha}{\nu_i, t}|^2$ is multiplied by the weight of the mass eigenstate $i$ in the incoherent solar flux, $P_{\nu_e \rightarrow \nu_i}^\odot(E)$.


The probability of oscillation for each flavour eigenstate in eq. \eqref{eq:sun_earth_probability} is implemented in \peanuts by the function \py{Pearth}, with signature
\begin{lstpy}
Pearth(nustate, density, pmns, DeltamSq21, DeltamSq3l, E, eta, depth, mode="analytical",
     massbasis=True, full_oscillation=False, antinu=False)
\end{lstpy}
which takes as arguments the neutrino state, \py{nustate}, an instance of the Earth density class (see Section \ref{sec:earthdensity} below), \py{density}, an instance of the PMNS class, \py{pmns}, the mass splitting parameters, \py{DeltamSq21} and \py{DeltamSq3l}, the neutrino energy, \py{E}, the nadir angle of the incoming neutrino, \py{eta}, and the depth of the experiment at which the probability is to be computed, \py{depth}. 

The optional argument \py{massbasis} defines the basis of the neutrino state. If \py{massbasis=False} the initial state is assumed to be a \emph{coherent} one, expressed in flavour basis, with \py{nustate} defining the complex coefficients $c_\alpha$ in eq.~(\ref{eq:general_flavour_state}), and final probabilities computed by squaring the coefficients evolved as in eq.~(\ref{eq:evolved_flavour_coeff}). If \py{massbasis=True}, the initial state is assumed to be an \emph{incoherent} superposition of mass eigenstates, with \py{nustate} defining the real weights $P_{\nu_e \rightarrow \nu_i}^\odot(E)$ and final probabilities computed as in eq.~(\ref{eq:sun_earth_probability}). To compute the probability for an antineutrino, one can set the optional argument \py{antinu=True} (\py{False} by default).

One can also optionally select the evolution mode to be either numerical or analytical (default) by providing the optional argument \py{mode} with either option. The function \py{Pearth} thus splits into two functions for each of the methods, \py{Pearth_numerical} and \py{Pearth_analytical}, whose details and differences will be described below in Section \ref{sec:propag}. Lastly, one can request the full evolution of the probability with the optional argument \py{full_oscillation} (defaults to \py{False}), returned as a list of probability values for each flavour eigenstate and certain discrete coordinate locations along the path of the neutrino. Note that the full oscillation can only be provided via the numerical evolution mode, so if selected along with the analytical mode, \peanuts will provide a warning and simply compute the final probabilities. 

In addition, if one wishes to know the final evolved complex coefficients $c_\alpha(t)$ from a coherent (flavour basis) initial neutrino state, from eq.~\ref{eq:evolved_flavour_coeff}, they can be obtained with the function
\begin{lstpy}
evolved_state(nustate, density, pmns, DeltamSq21, DeltamSq3l, E, eta, depth, mode="analytical", full_oscillation=False, antinu=False)
\end{lstpy}
which has the same arguments as \py{Pearth}, except the basis as this is only available for neutrino states in the flavour basis. As with the probability function, this function splits into \py{evolved_state_numerical} and \py{evolved_state_analytical} according to the selected mode.

\subsection{Evolutor operator}

The evolutor operator $\hat{\mathcal{U}}$ is defined by the equations
\begin{equation}
	\ket{\nu, t} = \hat{\mathcal{U}}(t, t_0) \ket{\nu, t_0}, \hspace{1cm} \text{ with } \hspace{1cm} \hat{\mathcal{U}}(t_0, t_0) = \hat{\mathbb{1}}.
\end{equation}
The equation of motion for $\hat{\mathcal{U}}(t, t_0)$ can be derived from the Schr{\"o}dinger equation
\begin{equation}
    i \frac{\de}{\de t}\ket{\nu, t} = \hat{H}(t) \ket{\nu, t} \Rightarrow i \frac{\de}{\de t} \hat{\mathcal{U}}(t, t_0) \ket{\nu, t_0}  = \hat{H}(t) \hat{\mathcal{U}}(t, t_0) \ket{\nu, t_0}, 
\end{equation}
implying
\begin{equation}\label{eq:evolutor_hamiltonian}
    i \frac{\de}{\de t} \hat{\mathcal{U}}(t, t_0)  = \hat{H}(t) \hat{\mathcal{U}}(t, t_0) \hspace{1cm} \text{ with } \hspace{1cm} \hat{\mathcal{U}}(t_0, t_0) = \hat{\mathbb{1}}.
\end{equation}

The Schr{\"o}dinger equation in flavour basis takes the form
\begin{equation}
    i \frac{\de}{\de t}\braket{\nu_\alpha}{\nu, t} = \bra{\nu_\alpha} \hat{H}(t) \ket{\nu_\beta} \braket{\nu_\beta}{\nu, t} \Rightarrow i \frac{\de}{\de t} c_\alpha(t) = H_{\alpha \beta}(t) c_\beta(t).
\end{equation}
The explicit expression for $H_{\alpha \beta}(t)$ can be readily derived in vacuum (where there is no time dependence)
\begin{equation}\label{eq:Ham_matrix}
    H_{\alpha \beta} = \bra{\nu_\alpha} \hat{H} \ket{\nu_\beta} = \bra{\nu_i} U_{\alpha i} \hat{H} U_{\beta j}^* \ket{\nu_j} = U_{\alpha i} U_{\beta j}^* \bra{\nu_i} \hat{H} \ket{\nu_j} = U_{\alpha i} U_{\beta j}^* E_j \delta_{ij} = \left[U \text{diag}(E_i) U^\dagger \right]_{\alpha \beta},
\end{equation}
where $E_i$ is the energy of the mass eigenstate $\ket{\nu_i}$.
In the presence of matter the Hamiltonian matrix elements receive an additional (time dependent) term, cf. Section~\ref{sec:propag}; Eq.~(\ref{eq:Ham_matrix}) determines its vacuum CP-structure, following the definition of the neutrino fields/states in Eq.s~(\ref{eq:nu_field_mixing}, \ref{eq:nu_state_mixing}).

In \peanuts the probability of oscillation through vacuum, and its evolved state, are computed with the functions
\begin{lstpy}
    Pvacuum(nustate, pmns, DeltamSq21, DeltamSq3l, E, L, antinu=False, massbasis=True)
    vacuum_evolved_state(nustate, pmns, DeltamSq21, DeltamSq3l, E, L, antinu=False)
\end{lstpy}
with similar arguments as \py{Pearth} above, with the notable difference of the oscillation length or baseline \py{L}, to be provided in km.


Once the Hamiltonian matrix elements are known, they can be used to derive the evolutor ones. The formal solution is
\begin{equation}
	\mathcal{U}(t, t_0) = \mathcal{T} \left[ e^{-i \int_{t_0}^t \de t' H(t')} \right],\label{eq:evolutor_formal}
\end{equation}
where $\mathcal{T}$ is the time-order operator. Eq.~(\ref{eq:evolutor_formal}) does not generally admit an analytic closed form, except for very special cases, for instance if the Hamiltonian at different times does commute.

A well known approach to the problem is the Dyson series~\cite{Dyson:1949bp} 
\begin{equation}
	\mathcal{U}(t, t_0) = \mathbb{1} + \sum_{n=1}^\infty \frac{\left(-i\right)^n}{n!} \int_{t_0}^t \de t_1 \int_{t_0}^t \de t_2 \cdots \int_{t_0}^t \de t_n \mathcal{T} \left[H(t_1) H(t_2) \cdots H(t_n) \right], \label{eq:Dyson_series}
\end{equation}
which allows for an approximate solution obtained by truncating Eq.~(\ref{eq:Dyson_series}) at finite values of $n$, if the series is expected to be perturbative.  We will return to this approximation in Section \ref{sec:propag}, where we will derive an approximated analytical expression for the evolutor.

%
%

\subsection{Earth matter regeneration}
\label{sec:earthdensity}

Due to the incoherent nature of the solar neutrino flux, its relative weights in mass eigenstates components, $P_{\nu_e \rightarrow \nu_i}^\odot(E)$, are constant while travelling from Sun to Earth in vacuum. However, before arriving at a terrestrial detector, a solar neutrino can cross the Earth itself, along which path the matter potential makes the propagation eigenstates different from the vacuum ones. This results in coherent neutrino oscillations inside the Earth that, on average, result in a regeneration of $\nu_e$ with respect to the vacuum case~\cite{Mikheev:1987wa}.

The electron density inside Earth can be parametrised by 5 shells, within which the density itself varies smoothly as~\cite{Lisi:1997yc}
\begin{equation}\label{eq:earth_density_param}
	N_j(r) = \alpha_j + \beta_j r^2 + \gamma_j r^4, \hspace{1cm} \text{with} \hspace{1cm} \left[N\right] = \text{mol}/\cm^3,
\end{equation}
and where $r$ is the radial distance normalised to the Earth radius. The numerical values of the parameters are reported for convenience in Table~\ref{tab:shell_parameters}.

\begin{table}[ht]
\begin{center}
\begin{tabular}{clcccc}
\hline
\hline
$j$& Shell          & $[r_{j-1},\,r_j]$  &$\alpha_j$ &$\beta_j$ &$\gamma_j$\\
\hline
1 & Inner core      &     $[0,\,0.192]$    &   6.099 & $-$4.119 &    0.000 \\
2 & Outer core      &   $[0.192,\,0.546]$  &   5.803 & $-$3.653 & $-$1.086 \\
3 & Lower mantle    &   $[0.546,\,0.895]$  &   3.156 & $-$1.459 &    0.280 \\
4 & Transition Zone &   $[0.895,\,0.937]$  &$-$5.376 &   19.210 &$-$12.520 \\
5 & Upper mantle    &     $[0.937,\,1]$    &  11.540 &$-$20.280 &   10.410 \\
\hline 
\hline
\end{tabular}
\caption{\label{tab:shell_parameters}Values of the parameters for the electron density expressed as $N_j(r) = \alpha_j + \beta_j r^2 + \gamma_j r^4$ with $[N] = \text{mol}/\cm^3$, for each of the Earth internal shells, as derived in~\cite{Lisi:1997yc}. The radial distance $r$ is normalised to the radius of Earth.}
\end{center}
\end{table}

\begin{figure}[ht]
	\includegraphics[width=\textwidth]{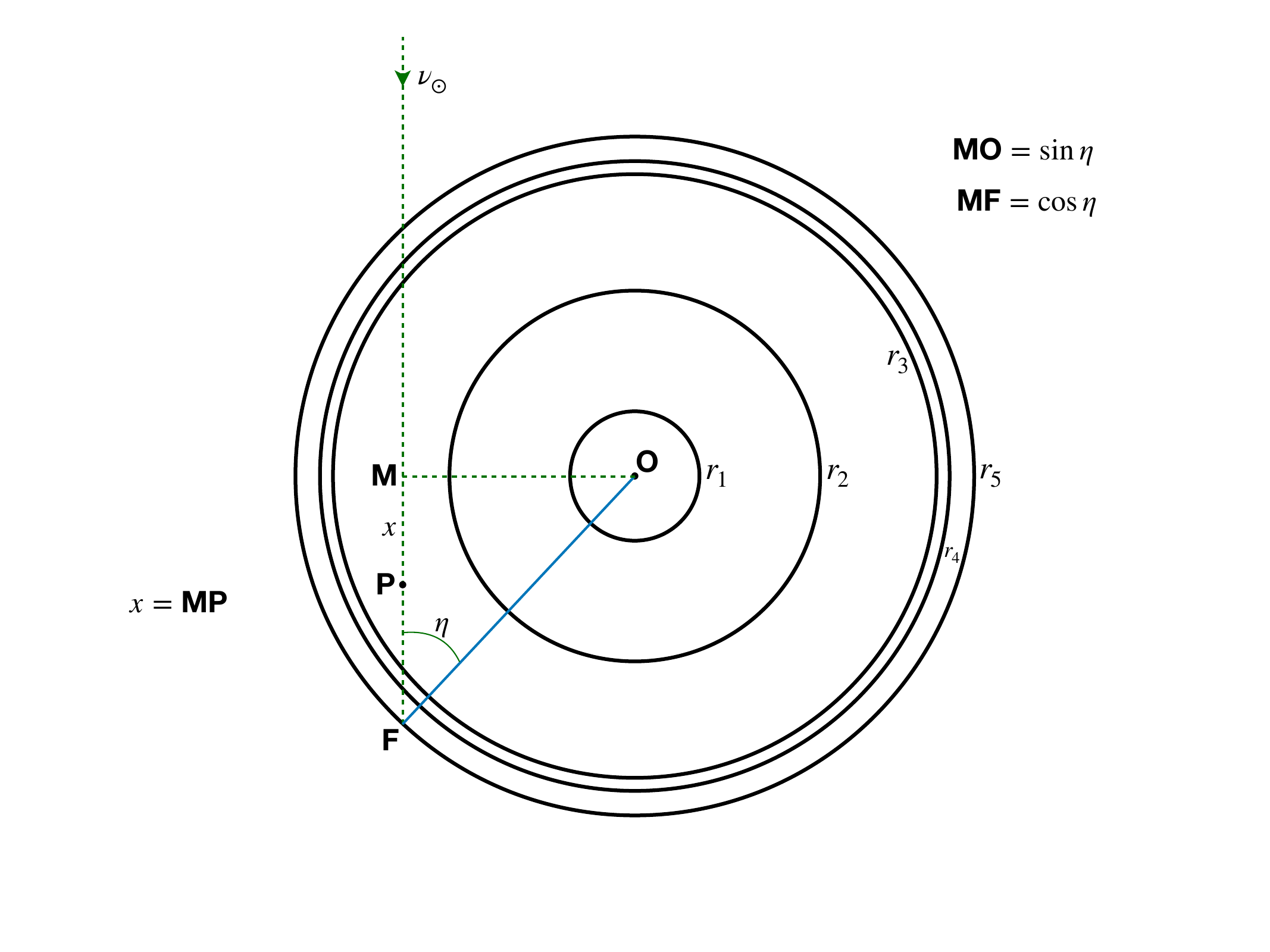}
	\caption{Earth section showing the different shells and the path of a solar neutrino having nadir angle $\eta$. The trajectory coordinate $x$ is also schematised.}
	\label{fig:earth_01}	
\end{figure}

\begin{figure}[ht]
\centering
\includegraphics[width=0.8\textwidth]{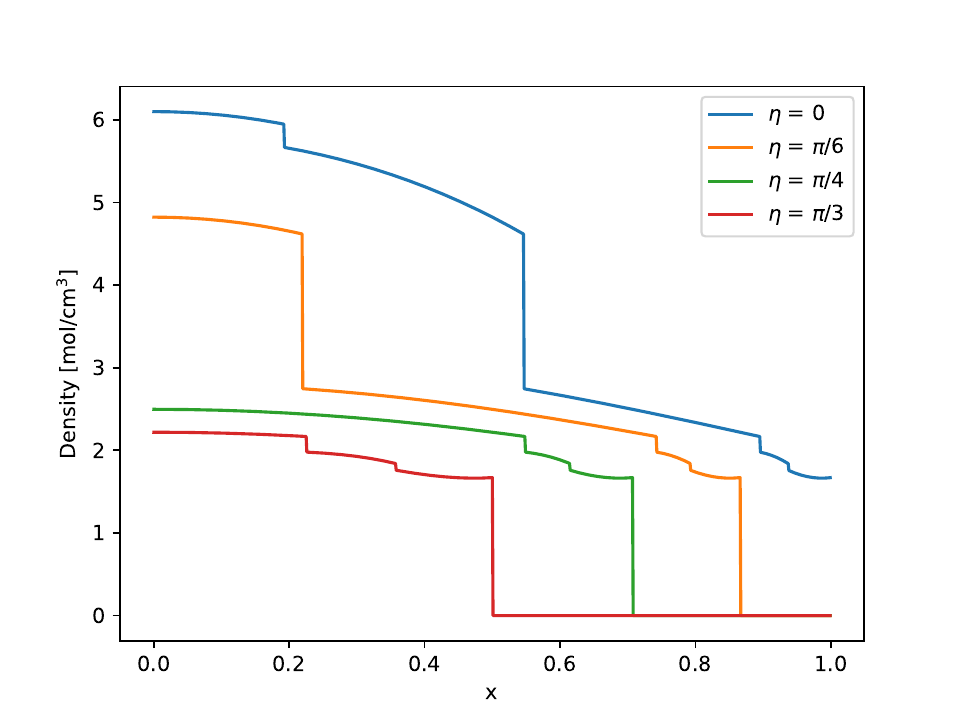}
\caption{Earth density profile for different values of the nadir angle $\eta$, following the parametrisation in~\cite{Lisi:1997yc}.}
\label{fig:earth_density}
\end{figure}

The parametrisation in Eq.~(\ref{eq:earth_density_param}) is valid for radial trajectories, i.e. paths crossing the center of the Earth. For a path forming a nadir angle $\eta$ with the radial trajectory (cf. Fig.~\ref{fig:earth_01}),
the parametrisation is functionally invariant, with modified coefficients 
\begin{eqnarray}
	N_j(x) &=& \alpha'_j + \beta'_j x^2 + \gamma'_j x^4, \label{eq:earth_density_nadir} \\
	\alpha'_j &=& \alpha_j + \beta_j \sin^2 \eta + \gamma_j \sin^4\eta, \\
	\beta'_j &=& \beta_j + 2 \gamma_j \sin^2\eta, \\
	\gamma'_j &=& \gamma_j,
\end{eqnarray}
where the \emph{trajectory coordinate} $x$ is defined as the distance from the trajectory mid-point, i.e. $x = \sqrt{r^2-\sin^2\eta}$. The Earth density profiles for some example values of the nadir angle $\eta$ are reported in Fig.~\ref{fig:earth_density}.

\begin{figure}[ht]
	\includegraphics[width=0.8\textwidth]{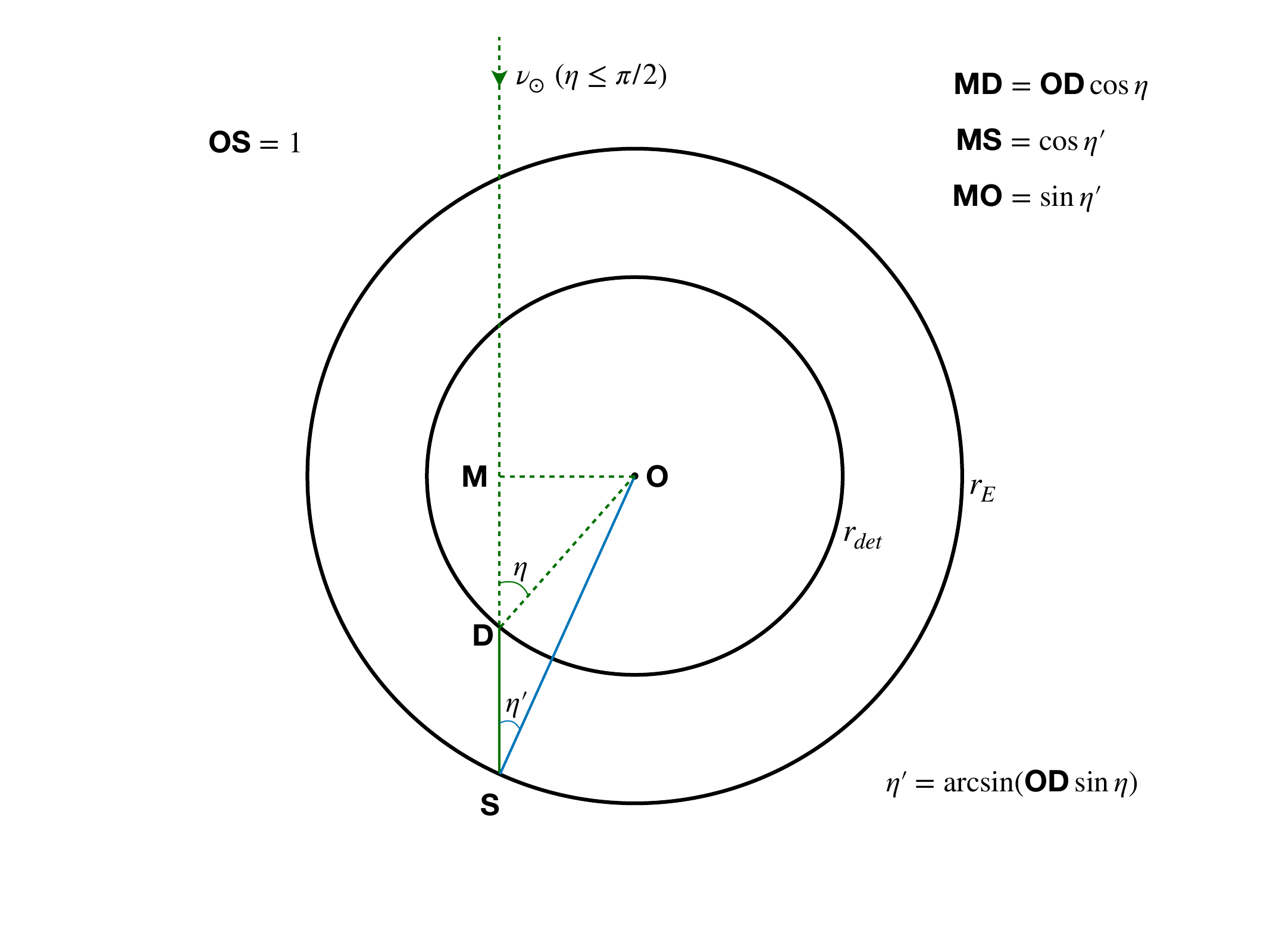}
	\caption{Schematic representation of Earth shell and ``detector shell'' for an underground detector.}
	\label{fig:scheme_detector_shell}
\end{figure}

This parametrisation assumes a nadir angle defined for a detection at the surface of Earth. If the detector is located underground, we can define a ``detector shell'' at the detector radial distance. For instance, SNO was placed $H=2$ km underground, so we can define $r_{det} = r_\text{SNO} = 1 - H /R_\text{E} \equiv 1 - h$, where $R_\text{E}$ is the (not rescaled) Earth radius, $R_E = 6.371 \cdot 10^3$ km. There are two modifications for the scenario of an underground detector with respect to the treatment in~\cite{Lisi:1997yc}: the first is that the measured nadir angle  $\eta$ differs from the angle $\eta' = \arcsin (r_{det} \sin \eta)$ that one would  measure at the Earth surface, for same neutrino trajectory, cf. Figure~(\ref{fig:scheme_detector_shell}). This implies that it is the $\eta'$ angle that should be used in Eq.~(\ref{eq:earth_density_nadir}) to compute the value of the electron density profile along the neutrino trajectory, and not the value $\eta$ measured by the experiment. The second modification is that matter effects are present even for values of $\eta \ge \pi/2$, if the detector is underground. 
The contribution to the trajectory from the outer layer (between Earth surface and detector shell) is given by
\begin{equation}
	 \Delta x(\eta) = \left\{ \begin{array}{l c c}
		-r_{det} \cos\eta + \sqrt{1 - r_{det}^2 \sin^2 \eta} & \text{for} & 0 \leq \eta \leq \frac{\pi}{2}, \\
		r_{det} \cos\eta + \sqrt{1 - r_{det}^2 \sin^2 \eta} & \text{for} & \frac{\pi}{2} \leq \eta \leq \pi.
	\end{array} \right.
\end{equation}
$\Delta x$ reduces to $h$ for $\eta=0, \pi$, and attains the maximum value $\sqrt{h(2-h)}$ for $\eta=\pi/2$ (this is approximately 160 km for SNO).

In general, not every shell is crossed by solar neutrinos: a shell $i$ is crossed by a neutrino trajectory with nadir angle $\eta$ if $r_i > r_{det} \sin \eta$. The value of the trajectory coordinate at each shell crossing is given by 
\begin{equation}
	x_i = \sqrt{r_i^2 - r_{det}^2 \sin^2 \eta}, \hspace{1cm} \text{ for $i$ such that} \hspace{1cm} r_i > r_{det} \sin \eta.
\end{equation}

\peanuts implements the Earth density as a \Python class called \py{EarthDensity}, with signature

\begin{lstpy}
    EarthDensity(density_file=None, tabulated_density=False, custom_density=False)
\end{lstpy}

where the optional arguments \py{density_file}, \py{tabulated_density}  and \py{custom_density} allow the user to supply any Earth density profile. If supplied by file, using the \py{density_file} argument, the density profile is either expected as a table with columns $\{r_j, \alpha_j, \beta_j, \gamma_j\}$, following eq.~\eqref{eq:earth_density_param} (additional optional columns $\delta_j^{2n}$, with $n\geq3$, are allowed, corresponding to higher orders in the polynomial expansion of the density); or if \py{tabulated_density=True} it is expected as a two-column table of radii and density, where each entry will be treated as a new layer of constant density. Consequently \peanuts can work with an arbitrary number of shells and arbitrary densities. In addition to supplying the density profile by file, it is also possible to provide an analytical expression for the density. Upon selecting the optional argument \py{custom_density=True} in the \py{EarthDensity} constructor, \peanuts will use the earth density computed by the member function \py{custom_density(r)}, where a user can implement their own analytical density profile. However, custom density profiles and higher orders in the polynomial expansion are only fully used when computing the oscillation probabilities numerically; the analytical computation, described below in Section \ref{sec:perturb}, relies on the density described as in eq.~\eqref{eq:earth_density_param}, and thus any density provided will be Taylor expanded and truncated to fit that form, when used for the analytical computation of the oscillation probability. Regardless of the source and form of the density profile, the \py{EarthDensity} class provides methods to compute the value of the density at given coordinate $x$, as well as, when appropriate, the modified coefficients $\alpha'_j$,  $\beta'_j$ and $\gamma'_j$ (and $\delta^{'2n}_j$ if needed), corresponding to each Earth shell, from the radii and nadir angle $\eta$ using Eq. \eqref{eq:earth_density_nadir} (or modified to accommodate higher orders).

%

\section{Neutrino propagation Hamiltonian}
\label{sec:propag}

The propagation Hamiltonian for an ultrarelativistic neutrino propagating in a medium with electron density $n_e(x)$ is, in the flavour basis~\cite{Fantini:2018itu}
\begin{equation}\label{eq:hamiltonian}
	H_\nu = \underbrace{U \text{diag}(k) U^\dagger}_{H_\nu^0} + \underbrace{V(x)\ \text{diag}(1,0,0)}_{V_\nu},
\end{equation}
with
\begin{eqnarray}
	k_i &=& \frac{m_i^2}{2E},\\
	V(x) &=& \sqrt{2}G_F n_e(x).
\end{eqnarray}
For antineutrinos, the same Eq.~(\ref{eq:hamiltonian}) holds, with the replacements
\begin{eqnarray}
	U &\rightarrow U^*,\\
	V &\rightarrow -V.
\end{eqnarray}

To streamline computations and statistical analysis, it is more convenient to redefine the Hamiltonian by subtracting a constant term

\begin{equation}
	H_\nu \rightarrow H_\nu - U \left(k_j \mathbb{1}\right) U^\dagger,
\end{equation}
with $j=1$ for normal ordering (NO) and $j=2$ for inverted ordering (IO), such that
\begin{eqnarray}
	\left(\begin{array}{ccc}
k_1 &&\\
& k_2 & \\
&& k_3
\end{array} \right) - k_1 \mathbb{1} &=& \frac{1}{2E} \left( \begin{array}{ccc}
 0 && \\
 & \Delta m_{21}^2 & \\
 && \Delta m_{31}^2
 \end{array} \right) \hspace{0.2cm} \text{ for NO}, \label{eq:kinetic_NO}\\
	\left(\begin{array}{ccc}
k_1 &&\\
& k_2 & \\
&& k_3
\end{array} \right) - k_2 \mathbb{1} &=& \frac{1}{2E} \left( \begin{array}{ccc}
 - \Delta m_{21}^2 && \\
 & 0 & \\
 && \Delta m_{32}^2
 \end{array} \right) \hspace{0.2cm} \text{ for IO}, \label{eq:kinetic_IO}
\end{eqnarray}
so that we can use as free parameters $\Delta m_{21}^2 >0$ and $\Delta m_{3\ell}^2$, with $\ell=1$ for NO ($\Delta m_{31}^2 > 0$) and $\ell=2$ for IO ($\Delta m_{32}^2 < 0$). In the following we keep using the notation $\text{diag}(k)$ for the general expressions, valid for any choice of mass ordering. A specific scenario can then be easily recovered by specifying the structure of this diagonal matrix, e.g. in Eq.s~(\ref{eq:kinetic_NO}, \ref{eq:kinetic_IO}). 

With the parametrisation in Eq.~(\ref{eq:pmns}) and using $\left[\Delta,R_{12}\right] = \left[V_\nu,R_{23}\right]=\left[V_\nu,\Delta\right]=0$, the propagation Hamiltonian can be rewritten as
\begin{equation}
	H_\nu = R_{23} \Delta \tilde{H} \Delta^* R_{23}^T,
\end{equation}
with
\begin{equation}
	\tilde{H} = R_{13} R_{12} \text{diag}(k) R_{12}^T R_{13}^T + V(x) \text{diag}(1,0,0).
\end{equation}
Notice that $\tilde{H}$ does not depend on $\theta_{23}$, $\delta$.

Given that $R_{23}$ and $\Delta$ do not depend on position, they can be factorised in the time-ordered definition of the evolutor operator
\begin{eqnarray}
	\mathcal{U} &=& \mathcal{T} e^{-i \int \de x H_\nu (x)} = \mathbb{1} + R_{23} \Delta \mathcal{T}\left[-i \int \de x \tilde{H}_\nu (x)\right] \Delta^* R_{23}^T +\\
	&& \frac{1}{2} R_{23} \Delta \mathcal{T}\left[ \left(\left(-i\right)^2 \int \de x_1 \de x_2 \tilde{H}_\nu (x_1) \underbrace{\Delta^* R_{23}^T R_{23} \Delta}_{= \mathbb{1}}  \tilde{H}_\nu (x_2)\right)\right] \Delta^* R_{23}^T \\
	&&  + \dots (\text{ other terms of the Dyson series for } n \rightarrow \infty ) \\
	&=& R_{23} \Delta \mathcal{T} \left[ e^{- i \int \de x \tilde{H}(x)} \right] \Delta^* R_{23}^T = R_{23} \Delta \tilde{\mathcal{U}} \Delta^* R_{23}^T. \label{eq:evolutor_factor}
\end{eqnarray}

A numerical solution for the evolutor can be obtained by solving eq. \eqref{eq:evolutor_factor} or by resolving the differential equation in eq. \eqref{eq:evolutor_hamiltonian}. \peanuts offers a numerical evaluation of the evolutor, via the function
\begin{lstpy}
Pearth_numerical(nustate, density, pmns, DeltamSq21, DeltamSq3l, E, eta, depth, massbasis=True, full_oscillation=False, antinu=False)
\end{lstpy}
and the evolved coefficients from a coherent neutrino state, with
\begin{lstpy}
evolved_state_numerical(nustate, density, pmns, DeltamSq21, DeltamSq3l, E, eta, depth, full_oscillation=False, antinu=False)
\end{lstpy}
These computations, however, can be extremely time-consuming, and thus it is convenient to find an approximated analytical expression by performing a perturbative expansion of the Hamiltonian.

\subsection{Perturbative expansion of the neutrino propagation Hamiltonian}
\label{sec:perturb}

We are interested in an expression for the operator in Eq.~(\ref{eq:evolutor_factor})
\begin{equation}\label{eq:evolutor_12-13}
	\tilde{\mathcal{U}} = \mathcal{T} \left[ e^{- i \int \de l \tilde{H}(l)} \right],
\end{equation}
where $l$ is the coordinate along the neutrino path. We normalise distances to the Earth radius $R_E$, by defining $x = l/R_E$. The Hamiltonian $\tilde{H}$ can be divided in a kinetic and a matter dependent terms
\begin{equation}
	\tilde{H}(x) = \tilde{H}_k + \sqrt{2} G_F n_e(x)\text{diag}(1,0,0),
\end{equation}
where $\tilde{H}_k$ does not depend on $x$.

To work out a perturbative expression for $\mathcal{U}$\footnote{For now we drop the tilde from the ``reduced'' evolutor $\tilde{\mathcal{U}}$ for covenience, and to avoid confusion with the average $\bar{\mathcal{U}}$. We will recover the notation at the end of the section when computing the full evolutor.} it is convenient to express the electron density as a perturbation along its mean value along the path~\cite{Lisi:1997yc}
\begin{equation}
	n_e(x) = \bar{n}_e + \delta n(x), \hspace{1cm} \bar{n}_e =\frac{1}{x_2-x_1} \int_{x_1}^{x_2} \de x \ n_e(x),
\end{equation}
from which it follows
\begin{equation}
	\int_{x_1}^{x_2} \de x\ \delta n(x) = 0.
\end{equation} 
We can analogously divide the Hamiltonian into a zeroth order term and a perturbation
\begin{equation}
	\tilde{H}(x) = \underbrace{\tilde{H}_k + \sqrt{2}G_F \bar{n}_e \text{diag}(1,0,0)}_{\tilde{H}_0} + \underbrace{\sqrt{2}G_F \delta n(x) \text{diag}(1,0,0)}_{\delta \tilde{H}(x)},
\end{equation}
where again $\tilde{H}_0$ does not depend on $x$.

The evolutor can thus be expressed as~\cite{Lisi:1997yc}
\begin{equation}
	\mathcal{U}(x_2,x_1) = \bar{\mathcal{U}}(x_2,x_1) - i \int_{x_1}^{x_2} \de x\ \bar{\mathcal{U}}(x_2,x)\ \delta \tilde{H}(x)\ \bar{\mathcal{U}}(x,x_1) + \mathcal{O}(\delta\tilde{H}^2),\label{eq:evolutor_1}
\end{equation}
where $\bar{\mathcal{U}}$ is the evolutor for constant matter density $\bar{n}_e$.

The evolutor for a constant Hamiltonian $\bar{H}$ can generally be expressed in a closed form~\cite{Ohlsson:1999xb}
\begin{equation}
	e^{- i \bar{H} x} = \phi \sum_{a=1}^3 e^{-i x \lambda_a} \frac{1}{3\lambda_a^2 + c_1}\left[ \left(\lambda_a^2 + c_1\right) \mathbb{1} + \lambda_a T + T^2 \right] \equiv \phi \sum_{a=1}^3 e^{-i x \lambda_a} M_a,
\end{equation}
where $T = \bar{H} - \text{Tr}(\bar{H}) \mathbb{1}/3 $ is a traceless matrix and $\lambda_a$ are the roots of the characteristic equation
\begin{equation}
	\lambda^3 + c_1 \lambda + c_0 = 0,
\end{equation}
with
\begin{eqnarray}
	c_1 &=& T_{11} T_{22} - T_{12} T_{21} + T_{11} T_{33} - T_{13} T_{31}
+ T_{22} T_{33} - T_{23} T_{32}, \\
c_0 &=& - \det T.
\end{eqnarray}
Finally
\begin{equation}
	\phi = e^{- i x \frac{\text{Tr}(\bar{H})}{3} }.
\end{equation}


By noticing that the full dependence on $x$ in $e^{-i \bar{H} x}$ is now contained within the scalar functions $e^{- i \lambda_a x}$, the first order correction in Eq.~(\ref{eq:evolutor_1}) can be computed as
\begin{eqnarray}
\mathcal{U}^{(1)}(x_2,x_1) &=&	- i \int_{x_1}^{x_2} \de x\ \bar{\mathcal{U}}(x_2,x)\ \delta \tilde{H}(x)\ \bar{\mathcal{U}}(x,x_1) \\
&=& - i \sum_{a,b=1}^3 \int_{x_1}^{x_2} \de x e^{- i \tilde{\lambda}_a (x_2-x)} M_a \text{diag}\left(\sqrt{2} G_F \delta n(x),0,0\right) M_b e^{- i \tilde{\lambda}_b (x-x_1)}\\
&=& - i \sum_{a,b=1}^3 M_a \text{diag}\left(\sqrt{2}G_F I_{ab}(x_2, x_1), 0, 0\right) M_b,
\end{eqnarray}
where  $\tilde{\lambda}_a = \lambda_a + \text{Tr}(\bar{H})/3$ and we defined
\begin{equation}
	I_{ab}(x_2,x_1) = \int_{x_1}^{x_2}\de x\ e^{- i \tilde{\lambda}_a (x_2-x)}\ \delta n(x)\ e^{- i \tilde{\lambda}_b (x-x_1)}.
\end{equation}
For a path fully contained within one shell we can parametrise 
\begin{equation}
	\delta n (x) = \tilde{\alpha}' + \beta' x^2 + \gamma' x^4,
\end{equation}
where $\tilde{\alpha}' = \alpha' - \bar{n}_e$, implying that $I_{ab}(x_2,x_1)$ can be expressed analytically in closed form.

Summarising, we can perturbatively expand the evolutor operator as 
\begin{eqnarray}
\mathcal{U}(x_2,x_1) &=& \mathcal{U}^{(0)}(x_2,x_1) +\mathcal{U}^{(1)}(x_2,x_1) + \mathcal{O}(\delta \tilde{H}^2),\\
\mathcal{U}^{(0)}(x_2,x_1) &=& 	e^{- i \bar{H} (x_2-x_1)} = \phi \sum_{a=1}^3 e^{-i (x_2-x_1) \lambda_a} M_a, \\ 
\mathcal{U}^{(1)}(x_2,x_1) &=& - i \sum_{a,b=1}^3 M_a\ \text{diag}\left(\sqrt{2}G_F I_{ab}(x_2, x_1), 0, 0\right)\ M_b.
\end{eqnarray}

In \peanuts we implement these perturbative expressions to compute $\mathcal{U}$ in Eq.~(\ref{eq:evolutor_12-13}) at first order in perturbation theory, using a ``reduced'' mixing matrix $\tilde{U} = R_{13} R_{12}$, with he function
\begin{lstpy}
Upert(DeltamSq21, DeltamSq3l, E, x2, x1, a, b, c, antiNu)
\end{lstpy}
which depends on the neutrino mass differences squared \py{DeltamSq21} and \py{DeltamSq3l}, the PMNS matrix \py{pmns}, neutrino energy \py{E}, the start and end points of the shell along the neutrino path \py{x2} and \py{x1}, as well as the density parameters of the traversed shell, with \py{a} = $\alpha$, \py{b} = $\beta$ and \py{c} = $\gamma$. The flag \py{anitNu} labels whether the computation is to be done for a neutrino (\py{antiNu=False}) or antineutrino (\py{antiNu=True}).

The procedure outlined above allows to express the evolutor at 1st order in $\delta H$, for a path fully contained within one shell. In general, the full evolutor on a generic path $(x_1,x_2)$ can be expressed as a time-ordered product of evolutors along the same path
\begin{equation}\label{eq:evolutor_product}
	\mathcal{U}(x_2, x_1) = \mathcal{U}(x_2, x_i) \mathcal{U}(x_i, x_1),
\end{equation}
where $x_i$ is a generic point $x_1 < x_i <x_2$ contained on the original path. It can be shown that~\cite{Lisi:1997yc}
\begin{equation}\label{eq:evolutor_transpose}
	\mathcal{U}(0, -x) = \mathcal{U}(x,0)^T.
\end{equation}
The consequences of Eq.s~(\ref{eq:evolutor_product},~\ref{eq:evolutor_transpose}) are twofold: first, for a path starting at $x=x_i$, with $0\leq  x_i < x_1$,  crossing $n$ shells with boundaries at trajectory coordinate $(0, x_1, x_2, \dots, x_n)$, and ending at the point $x=x_f$, with $x_{n-1} < x_f \leq x_n$, the full evolutor can be expressed as
\begin{equation}
	\mathcal{U}(x_f, x_i) = \mathcal{U}(x_f, x_{n-1})\mathcal{U}(x_{n-1},x_{n-2})\dots \mathcal{U}(x_2, x_1)\mathcal{U}(x_1, x_i).
\end{equation}
Second, for detectors placed at surface, the Earth spherical symmetry implies that the electron density is symmetric with respect to the trajectory midpoint at $x=0$; thus we only need to compute the evolutor on one half-path (cf. Fig.~\ref{fig:earth_01})
\begin{equation}
	\mathcal{U}(x_F,x_I) = \mathcal{U}(x_F, 0) \mathcal{U}(0, - x_F) = \mathcal{U}(x_F, 0) \mathcal{U}(x_F, 0)^T.
\end{equation}
Notice that the final evolutor is only a function of $\eta$, since both the density profile and travelled distance within Earth are a function of the nadir angle.

If the detector is placed underground at trajectory coordinate $x_{det} < x_F$, we distinguish two cases. For $0 \leq \eta < \pi/2$
\begin{equation}
	\mathcal{U}(x_{det},x_I) = \mathcal{U}(x_{det}, 0) \mathcal{U}(0, - x_F) = \mathcal{U}(x_{det}, 0) \mathcal{U}(x_F, 0)^T, \hspace{0.2cm} \left(0 \leq \eta < \frac{\pi}{2} \right)
\end{equation}
where $x_{det} = r_{det} \cos{\eta}$. For $\pi/2 \leq \eta \leq \pi$ the electron density can be approximated to a constant value, since the neutrino path is never deeper than $H$ and density variations are negligible for realistic detectors. For instance, taking the SNO reference value $H = 2$ km,
\begin{equation}
	\frac{N(R_E) - N(R_E-H)}{N(R_E) + N(R_E - H)} = 10^{-4}.
\end{equation}
We fix for simplicity the electron density value to the one at Earth surface, $n_1 = 1.67$ mol/cm${^3}$. Having assumed constant density along the path, the evolutor is simply given by
\begin{equation}
	\mathcal{U}(\eta) = e^{-i R_E \Delta x(\eta) \left(\tilde{H}_0 + \text{diag}\left(\sqrt{2}G_F n_1, 0, 0\right) \right)}, \hspace{0.5cm} \left(\frac{\pi}{2} \leq \eta < \pi \right).
\end{equation}

To obtain the full dynamics and thus the full evolutor $\mathcal{U}$, \peanuts computes the time-ordered product of ``reduced'' evolutors $\tilde{\mathcal{U}}$ from above, and inputs it in Eq.~(\ref{eq:evolutor_factor}) to re-introduce the dependence on $\theta_{23}$ and $\delta$. The \peanuts function that implements these two steps is
\begin{lstpy}
FullEvolutor(density, DeltamSq21, DeltamSq3l, pmns, E, eta, depth, antiNu)
\end{lstpy}
with dependency also on the mass differences squared, PMNS matrix and neutrino energy, but also on the full Earth density profile \py{density}, the nadir angle \py{eta} and experiment depth \py{depth}. This full evolutor is finally used to compute the probability of oscillation at the detector by the function
\begin{lstpy}
Pearth_analytical(nustate, density, pmns, DeltamSq21, DeltamSq3l, E, eta, depth, massbasis=True, antinu=False)
\end{lstpy}
and the evolved coefficients with
\begin{lstpy}
evolved_state_analytical(nustate, density, pmns, DeltamSq21, DeltamSq3l, E, eta, depth, antinu=False)
\end{lstpy}

\section{Time integration and exposure}
\label{sec:time_integration}

Solar neutrino experiments typically collect data over a finite interval of time. As such, the measured survival probability is averaged over exposure
\begin{equation}\label{eq:time_integral_general}
	\left< P_E \right> = \frac{\int_{\tau_{d_1}}^{\tau_{d_2}} \de \tau_d \int_{\tau_{h_1} (\tau_d)}^{\tau_{h_2} (\tau_d)} \de \tau_h P_E \left(\eta(\tau_d, \tau_h) \right)}{\int_{\tau_{d_1}}^{\tau_{d_2}} \de \tau_d \int_{\tau_{h_1} (\tau_d)}^{\tau_{h_2} (\tau_d)} \de \tau_h },
\end{equation}

where $\tau_d$ is the daily time, $\tau_h$ the hourly time and $\eta$ the Sun nadir angle at detector location. The integration in Eq.~(\ref{eq:time_integral_general}) is typically not the most convenient choice for practical applications; a more effective option is to transform the double integral into a single one over $\eta$~\cite{Lisi:1997yc}
\begin{equation}\label{eq:eta_weight_integral}
	\left< P_E \right> = \int_{0}^{\pi} \de \eta W(\eta) P_E(\eta),
\end{equation}
where $W(\eta)$ is a normalized weight function representing the fraction of time in which the experiment collected data at nadir angle $\eta$.
For real experiments, $W(\eta)$ must be provided by the collaboration, taking into account the actual times at which the detector collected data or has been offline. It is nevertheless possible to compute $W(\eta)$ analytically, for the ideal case of an experiment continuously taking data between days $\tau_{d_1}$ and $\tau_{d_2}$~\cite{Lisi:1997yc}.
This is done by changing integration variables
\begin{eqnarray}
	\int_{\tau_{d_1}}^{\tau_{d_2}} \de \tau_d \int_{\tau_{h_1} (\tau_d)}^{\tau_{h_2} (\tau_d)} \de \tau_h P_E \left(\eta(\tau_d, \tau_h) \right) &=& \int_{\tau_{d_1}}^{\tau_{d_2}} \de \tau_d \int_{0}^{\pi} \de \eta \frac{\de \tau_h (\tau_d, \eta)}{\de \eta} P_E \left(\eta \right) \\
	= \int_{0}^{\pi} \de \eta  P_E \left(\eta \right) \int_{\tau_{d_1}}^{\tau_{d_2}} \de \tau_d \frac{\de \tau_h (\tau_d, \eta)}{\de \eta} &=& \int_{0}^{\pi} \de \eta P_E(\eta) W(\eta). \label{eq:integral_W}
\end{eqnarray}

By normalising the daily and hourly times to the interval $[0, 2 \pi]$
\begin{equation}
	\tau_d = \frac{\text{day}}{365} 2\pi, \hspace{2cm} \tau_h = \frac{\text{hour}}{24} 2\pi,
\end{equation}
with $\tau_d=0$ at winter solstice and $\tau_h=0$ at the middle of the night, it is possible to express
\begin{equation}
	\tau_h = \arccos\left(\frac{\sin(\lambda) \sin(\delta_S) + \cos(\eta)}{\cos(\lambda) \cos(\delta_S)} \right),
\end{equation}
with $\lambda$ the detector latitude and $\delta_S$ the Sun declination, given by
\begin{equation}
	\delta_S = \arcsin\left(-\sin(i) \cos(\tau_d)\right),
\end{equation}
with $i=0.4091$ rad being the Earth inclination. With these definitions it is possible to perform the integral defining $W(\eta)$ in  Eq.~(\ref{eq:integral_W}); it is convenient to restrict $\tau_d$ within the interval $[0, \pi]$ (the alternative case can be easily derived from this one by using the symmetry of the orbit) and to change the integration variable from $\tau_d$ to $T = \cos(\tau_d)$. 
The resulting indefinite integral is expressed in terms of elementary functions and of the incomplete elliptic integral of the first kind; its analytic expression is not particularly illuminating but can be easily evaluated numerically.
Some care must be taken in defining the range of integration for the definite integral, as this is given by the intersection of three distinct intervals: i) $T \in [-1, 1]$ is the interval where $T=\cos(\tau_d)$ is defined, ii) $T \in [\sin(\lambda - \eta)/\sin(i), \sin(\lambda + \eta)/\sin(i)]$ is the range where $T$ can take values for fixed values of $\eta, \lambda, i$, iii) $T \in [\cos(\tau_{d_2}), \cos(\tau_{d_1})]$ is the observation time. If the intersection of the three intervals is null then $W(\eta)$ will vanish for that given combination of $\lambda, \eta$ values.

The exposure function $W(\eta)$ is computed in \peanuts by the function

\begin{lstpy}
NadirExposure(lam=-1, d1=0, d2=365, ns=1000, normalized=False, from_file=None, angle="Nadir")
\end{lstpy}
which has no required arguments, but either the latitude of the experiment, \py{lam}, or an exposure file \py{from_file}, must be provided. It returns tabulated values of the function $W(\eta)$ for \py{ns} samples (default 1000) in nadir angle $\eta$, assuming an exposure from day \py{d1} to day \py{d2}, where \py{d1=0} corresponds to the northern hemisphere winter solstice. The exposure may be selected to be normalized with the option \py{normalized} (defaults to \py{False}). 

\begin{figure}[ht]
\centering
	\includegraphics[width=0.8\textwidth]{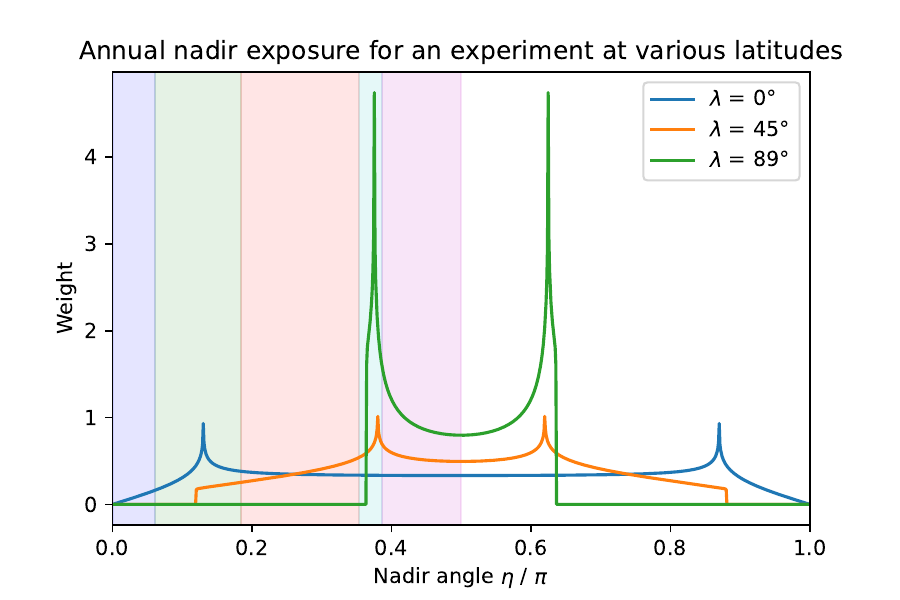}	
	\caption{Weight of the nadir angle exposure for an ideal experiment located at latitude $\lambda$, taking data continuously over a full year. The coloured regions represent the nadir angles subtending the Earth internal shells as parametrised in Table~\ref{tab:shell_parameters}, for a detector located at the Earth surface.}
	\label{fig:eta_weights_ideal}
\end{figure}

Under default conditions, the function \py{NadirExposure} computes analytically the ideal exposure function performing the integral in eq. \eqref{eq:integral_W}, which is implemented in \peanuts by the function
\begin{lstpy}
IntegralDay(eta, lam, d1=0, d2=365)
\end{lstpy}
which depends on the nadir angle \py{eta}, latitude \py{lam} and day interval \{\py{d1},\py{d2}\}. We plot in Fig.~\ref{fig:eta_weights_ideal} the exposure function $W(\eta)$ for one full year of exposure for three ideal detectors located at latitudes $\lambda = 0, 45^\circ, 89^\circ$.

In a realistic case, the exposure must be provided by the experiment. For this purpose, one can provide an exposure file via the option \py{from_file} of the function \py{NadirExposure}, which shall contain the tabulated values of the exposure. By default it is assumed that the exposure is tabulated in values of the nadir angle $\eta$, but in some cases, the experiments provide the exposure tabulated in either the zenith angle $\theta$ or $\cos\theta$. In those cases, one must specify which angle is used for the tabulation with the options \py{angle="Zenith"} or \py{angle="CosZenith"}, respectively. If an exposure file is provided, the latitude is no longer required, and if the argument \py{lam} is provided it will be ignored. Note that irrespective of the original tabulated values, the function \py{NadirExposure} only returns values tabulated in $\eta$.

Lastly, \peanuts provides the averaged probability of oscillation at detector location taking into account the finite time exposure, as in eq. \eqref{eq:eta_weight_integral}, with the function
\begin{lstpy}
Pearth_integrated(nustate, density, pmns, DeltamSq21, DeltamSq3l, E, depth, mode="analytical", full_oscillation=False, antinu=False, lam=-1, d1=0, d2=365, ns=1000, normalized=False, from_file=None, angle="Nadir", daynight=None)
\end{lstpy}
which has the same arguments as the \py{Pearth} function, but without the nadir angle \py{eta}, and adds the same optional arguments as \py{NadirExposure}, with the final addition of an optional argument  \py{daynight} to select integrating only the nadir angles corresponding to the night period $\eta < \pi/2$, with \py{daynight="night"}, or the day period $\eta \geq \pi/2$, with \py{daynight="day"}. Note that this function requires the input neutrino state to be on the mass basis, since the integration is performed for an incoming incoherent flux of mass eigenstates, and thus it is not possible to provide a state on the flavour basis.

\section{Quick start guide}
\label{sec:quickstart}

\peanuts is an open-source software written in \Python, and as such we expect it to run seamlessly in every \Python 3 environment, irrespective of the architecture. Nevertheless it has been tested thoroughly on Linux and Mac OS X systems. An example installation guide using the \texttt{conda} environment can be found in Appendix \ref{app:installation}. \peanuts is optimised to run fast with minimal I/O operations, in order to allow effortless interface with other frameworks. Hence, it uses extensively the pre-compile \texttt{numpy} package as well as just-in-time compilation from \texttt{numba}. The full list of required and optional packages is as follows
\begin{itemize}
 \item[-] \textsf{numpy}: fast pre-compiled array operations
 \item[-] \textsf{numba}: just-in-time compilation of functions and classes\footnote{The Python package \textsf{numba} is an experimental package and thus prone to errors. In spite of that, \textsf{numba} is required to run \peanuts, since computational speed is one of the main goals. As of the release version, \peanuts works smoothly with \textsf{numba} 0.56. Some \texttt{conda\xspace} environments have shown issues with versions $\geq 0.57$. In those cases, we recommend reverting back to 0.56, which should work effortlessly.}
 \item[-] \textsf{os}:  file paths
 \item[-] \textsf{copy}: shallow and deep copies of objects
 \item[-] \textsf{time}: timing information
 \item[-] \textsf{math}, \textsf{cmath}, \textsf{mpmath}: mathematical operations
 \item[-] \textsf{scipy}: integration and interpolation routines
 \item[-] \textsf{pyinterval}: define integration intervals
 \item[-] \textsf{decimal}: output formatting
 \item[-] \textsf{pandas}: reading of csv files
 \item[-] \textsf{pyyaml}: reading of yaml files \textit{Optional}
 \item[-] \textsf{pyslha}: reading of slha files \textit{Optional}
 \item[-] \textsf{gitpython}: extract git tags \textit{Optional}
\end{itemize}

There are two operational modes of \peanuts, which we will call the \textit{simple} and \textit{expert} modes. The \textit{simple} mode allows \peanuts to be run just from the command line, appending the input parameters as arguments to the command. Naturally this is a more limited mode as only provides a fraction of the functionalities, but it is a fast way to run \peanuts just from the command line. The \textit{expert} mode uses all of the functionalities in \peanuts, and thus requires a configuration file, in \yaml format, to be written.

In the \textit{simple} mode one can run one of two provided scripts \py{run_prob_sun.py} and \py{run_prob_earth.py}. The first script computes the probability of neutrinos at the surface of the Sun, and has a signature
\begin{lstpy}
 run_prob_sun.py [options] <energy> <fraction> [ <th12> <th13> <th23> <delta> <dm21> <dm3l>]
\end{lstpy}
where \py{<energy>} and \py{<fraction>} are mandatory arguments that refer to the neutrino energy, in MeV,  and fraction respectively. The list of available neutrino fractions is \py{["pp", "hep", "7Be", "8B", "13N", "15O", "17F"]}\footnote{The default $^7Be$ spectrum comes from  ground-state transitions, which can also be provided as \py{"7Beground"}. One can alternatively use the transitions from the excited $^7Be$ state with \py{"7Beexcited"}.}. With no options provided, the remaining arguments are required to populate the PMNS matrix and the mass splittings. However, if the option \py{-i/--in_slha <slha_file>} is provided, a file location is expected after the flag, corresponding to a SLHA file where the neutrino parameters are defined, and thus only two arguments are required. Note that this method only works if the package \textsf{pyslha} is installed. Other options are \py{-s <solar_model>} for a different solar model file, \py{-v/--verbose} for verbose output and \py{-h/--help} to print usage information.

The second script \py{run_prob_earth.py} computes the probability of neutrinos at a given experimental location below the Earth's surface. It is used in the following way
\begin{lstpy}
run_prob_earth.py [options] -f/-m <state> <energy> <eta> <depth> [<th12> <th13> <th23>  <delta> <dm21> <dm3l>]
\end{lstpy}
where, as before, one can provide the full PMNS parameters and mass splittings explicitly, or as an SLHA file (with the \py{-i/--in_slha <slha_file>} option). In addition one must provide the initial neutrino state with the option \py{ -f/--flavour <state>} in the flavour basis, or \py{-m/--mass <state>} in the mass basis for an incoherent incoming flux. The neutrino state must be given as comma-separated list of three real or complex numbers, without spaces in between them, e.g.
\begin{lstpy}
run_prob_earth.py -f 0.1+0.03j,0.6+0.05j,0.09-0.9j ...
\end{lstpy}
Other necessary arguments are the neutrino energy \py{<energy>}, in MeV, the nadir angle, \py{<eta>}, in radians, and the depth of the experiment \py{<depth>}, in meters. Additional options include \py{-d/--density <density>} to provide a different Earth density file, \py{--antinu} to perform the computations for antineutrinos, \py{--analytical}  or \py{--numerical} to select either analytical or numerical evolution, \py{-v/--verbose} for verbose output and \py{-h/--help} to print usage information.

The \textit{expert} mode allows the user to exploit all the functionalities of PEANUTS. It requires writing a configuration file in \yaml format, hence this mode is only available if the optional \texttt{pyyaml} module is installed. For a given \yaml file, the \textit{expert} mode can then be used by using the \py{run_peanuts.py} executable in the following way
\begin{lstpy}
run_peanuts.py -f <my_yaml_file>
\end{lstpy}

Since most of the options are provided in the \yaml file, this command only takes the options \py{-v/--verbose} for verbose output and \py{-h/--help} to print usage info. An example \yaml file to compute the probability at the surface of the Sun can be seen below

\begin{lstpy}
Energy: 15

Neutrinos:
  dm21: 7.42e-05
  dm3l: 2.51e-03
  theta12: 5.83638e-01
  theta23: 8.5521e-01
  theta13: 1.49575e-01
  delta: 3.40339

Solar:
  fraction: "hep"
  flux: true
  spectrum: "distorted"
\end{lstpy}

The \yaml file must contain the \py{Energy} and \py{Neutrinos} nodes, as well as either the \py{Vacuum}, \py{Solar} or \py{Earth} nodes. The \py{Solar} and \py{Earth} nodes can appear simultaneously to combine the effects, whereas the \py{Vacuum} node cannot be combined with either. The \py{Energy} node must simply contain a real number representing the neutrino energy in MeV. The \py{Neutrinos} node can either contain a map of the neutrino parameters, as in the example above, or a single string with the location of a SLHA file (provided \textsf{pyslha} is installed), always relative to the location of the \py{run_peanuts.py} executable, in the following way
\begin{lstpy}
Neutrinos: "examples/example_slha.slha2"
\end{lstpy}

The \py{Solar} node must contain a map with at least an entry for the neutrino fraction \py{fraction}. As above, the list of available neutrino fractions is \py{["pp", "hep", "7Be", "8B", "13N", "15O", "17F"]}. By default, this will compute and print the probabilities of oscillation at Sun exit for all flavour eigenstates, but it can be disabled with the entry \py{probabilities: false}. Additionally, one can choose to print the total flux with \py{flux:true} (defaults to \py{false}) and the distorted or undistorted spectrum with \py{spectrum:"distorted"} or \py{spectrum:"undistorted"}, both of which are disabled by default. Lastly one can select the specific solar model to use with the entry \py{solar_model}, the flux file with \py{flux_file} and the spectrum files with the entry \py{spectra}. Note that if the solar model is not known to \peanuts, in addition to the path to the relevant files, one must provide the location within those files where the information can be found, that is the entries \py{fluxrows}, \py{fluxcols}, \py{fluxscale}, \py{distrow}, \py{radiuscol}, \py{densitycol} and \py{fractioncols}, in the format described above in Section \ref{sec:solar}.

The \py{Earth} node must contain a map with necessary entries to compute the probability at some location on Earth. Consequently, it requires an entry for the depth under the Earth's surface, \py{depth}, and either a nadir angle \py{eta} or a \py{latitude}. In addition, if the \py{Solar} node is not present in the \yaml file, a neutrino state is required as the \py{state} entry (see below for an example), as well as an entry to specify in which basis it is, \py{basis: "flavour"} (coherent) or \py{basis: "mass"} (incoherent). If the entry \py{antinu: True} is select, the computations will be performed for an antineutrino. One can provide a user-defined Earth density profile by providing a density file with the entry \py{density}, and indicate whether the density comes from tabulated data, with the entry \py{tabulated_density} or from a custom analytical expression, with the entry \py{custom_density}. It is also possible to choose either the numerical or analytical computation of the evolutor with \py{evolution: "numerical"} or \py{evolution: "analytical"} (the latter being the default). Lastly, if the entry \py{latitude} is present, the probabilities will be computed integrated over exposure, and thus one provides entries to modify the normalization, days interval, number of samples, exposure file and exposure angle (see Section \ref{sec:time_integration} for the meaning of these quantities) with the options \py{exposure_normalized} (\py{True} or \py{False}), \py{exposure_time} ([\py{d1},\py{d2}]), \py{exposure_samples} (number of samples), \py{exposure_file} (path) and \py{exposure_angle} (\py{"Nadir"}, \py{"Zenith"} or \py{"CosZenith"}), respectively. An example \yaml file with only the earth node can be seen below

\begin{lstpy}
Energy: 20

Neutrinos:
  dm21: 7.42e-05
  dm3l: 2.51e-03
  theta12: 5.83638e-01
  theta23: 8.5521e-01
  theta13: 1.49575e-01
  delta: 3.40339

Earth:
  state: [0.4, 0.2, 0.6]
  basis: "flavour"
  eta: 0.8
  depth: 3000
\end{lstpy}

The \py{Vacuum} node must contain the minimal requirements to compute oscillations in vacuum. This means it requires an input neutrino \py{state} and its \py{basis}, as for the earth oscillations above. It must also contain the distance traveled as \py{baseline}, in km. Lastly, optionally one can request vacuum oscillations for antineutrinos with \py{antinu: True}, and to switch on or off the printing of the probabilities or the evolved state with the \py{probabilities} and \py{evolved_state} options, respectively.

By default, the results of a \peanuts run will be printed to screen, but one can redirect the ouput to a file by adding the \py{Output} node to the \yaml file. This node must contain a single string corresponding to the output file location, which will be created if it does not exist, or \py{"stdout"} to print to screen (default). As an example, in order to redirect output to a file called \texttt{out.dat} in the same directory as the executable, the following node must be added to the \yaml file
\begin{lstpy}
Output: "out.dat"
\end{lstpy}

Finally, \peanuts allows the possibility to run simple grid scans of the input parameters provided in the \yaml file. This can be achieved easily by providing a range of values instead of a single real value for the parameters. This range can be provided as \py{[min,max,step]}, where a specific step size is selected, or \py{[min,max]}, where the step will be computed so that a total of 10 samples for the parameter are produced. For instance, to scan over the energy between 20 and 100 MeV, with a step of 10 MeV, one could add the following to the \yaml file.

\begin{lstpy}
Energy: [20,100,10]
\end{lstpy}

\peanuts can perform such simplistic grid scans on the neutrino energy (as seen above), as well as on any neutrino parameter ($\Delta m_{21}^2$, $\Delta m_{3l}^2$, $\theta_{12}$, $\theta_{13}$, $\theta_{23}$ and $\delta$)\footnote{Although it is technically possible to specify a range for $\Delta m_{3l}^2$ that includes negative and positive values, we do not recommend doing so, as the meaning of parameter changes from negative to positive (as the ordering of mass eigenstates change), which is statistically inconsistent. Hence we recommend scanning negative and positive values separately.}, and on the nadir angle \py{eta}. Furthermore, one can scan a parameter in log scale by setting the parameter entry in the \yaml file as \py{[min,max,step,"log"]}, in which case \py{min}, \py{max} and \py{step} are taken to be in log scale too. As these are crude grid scans, we do not recommend using this functionality for thorough scans of the full neutrino parameter space. For that purpose one should use smart sampling algorithms (see \cite{AbdusSalam:2020rdj}), which can be easily interfaced with \peanuts\footnote{This was used, for instance, in an upcoming global study of neutrino oscillations by the \textsf{GAMBIT} neutrino working group\cite{GAMBITNus}.}.

\subsection{Validation}

We have validated the results from \peanuts with those published by the SNO experiment~\cite{FiuzadeBarros:2011qna,SNO:2006odc}, both against their predictions, e.g. fluxes and spectra, and their measured data, e.g. exposure and observed events.  Here we document some of the validation tests that were performed. These can be reproduced by running the \py{run_SNO_test.py} file shipped with \peanuts, in the following way
\begin{lstpy}
run_SNO_test.py [options]
\end{lstpy}
which allows for additional options to be supplied, such as \py{-s/--solar <file>} to add a custom solar model, \py{-d/--density <file>} to add a custom earth density profile, \py{-f/--file <file>} to modify the default neutrino parameters with an SLHA file, and the ever present options \py{-v/--verbose} for debug output and \py{-h/--help} for usage info.

In order to compare the probability results with those from SNO, we use the same input values for the neutrino parameters, $\theta_{12}$, $\theta_{13}$, $\Delta m_{21}^2$ and $\Delta m_{3l}^2 = \Delta m_{31}^2$. The values of $\theta_{23}$ and $\delta$ are, for the most part, irrelevant for this comparison, but required by \peanuts to build the full PMNS matrix, so for those we take the values found in \cite{Esteban:2020cvm}. Thus, the values used here are
\begin{align}
\tan^2\theta_{12} = 0.469, \quad \sin^2 \theta_{13} = 0.01, \quad \theta_{23} = 0.85521, \quad \delta = 3.4034\notag \\
 \Delta m_{21}^2 = 7.9\times 10^{-5}, \quad \Delta m_{31}^2 = 2.46\times10^{-3}.
\end{align}

\begin{figure}[ht]
 \includegraphics[width=0.5\textwidth]{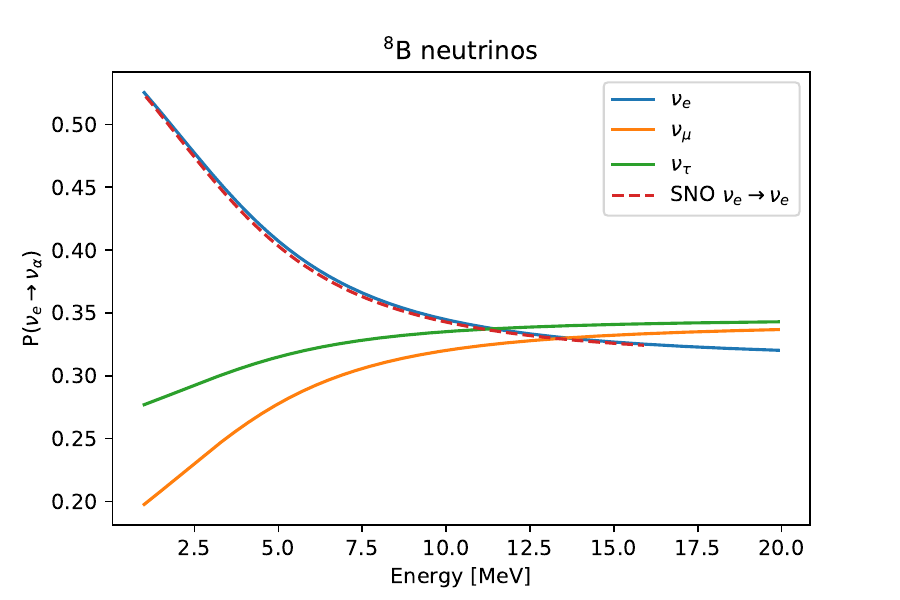}
 \includegraphics[width=0.5\textwidth]{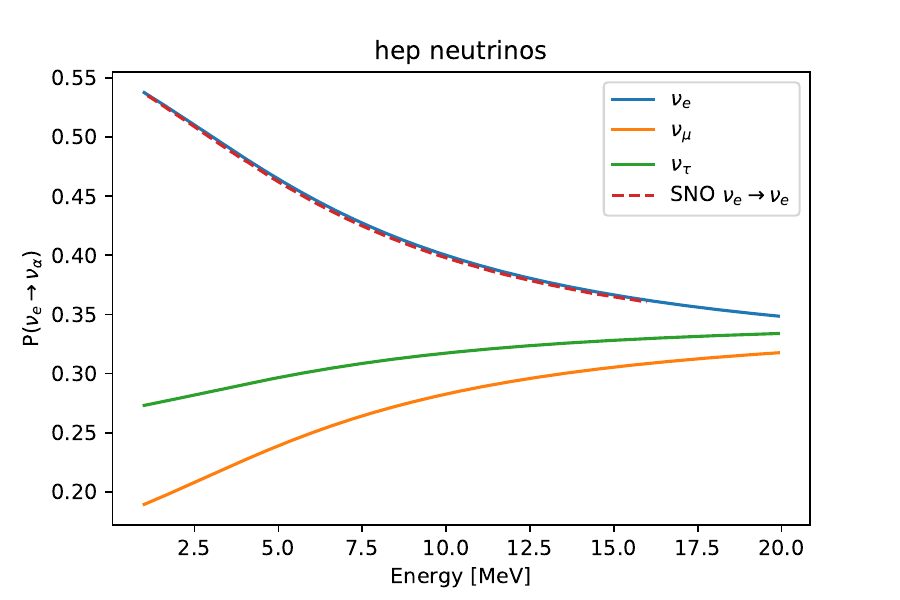}
\caption{Survival probability at the surface of the Sun for $^8B$ (left) and $hep$ (right) neutrinos. The solid lines are the \peanuts predictions, the dashed lines are the digitised curves for the SNO experiment, from Fig. 6.3 in~\cite{FiuzadeBarros:2011qna}.}
\label{fig:SNO_8B_hep_comparison}
\end{figure}

With these values, we can compare in Fig.~\ref{fig:SNO_8B_hep_comparison} the prediction from our code with the SNO survival probabilities at the surface of the Sun for the ${}^8B$ and $hep$ neutrinos, digitised from Fig. 6.3 in~\cite{FiuzadeBarros:2011qna}. Naturally, for the same Solar model, the predictions of \peanuts for each neutrino fraction match very well that reported by the SNO experiment.

\begin{figure}[ht]
 \includegraphics[width=0.5\textwidth]{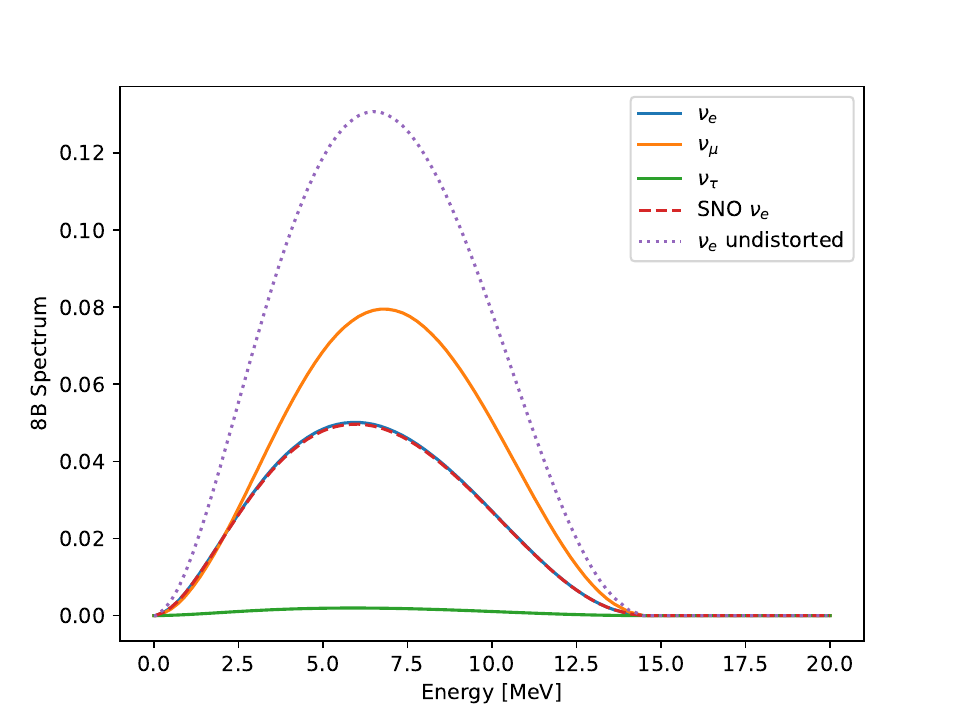}
 \includegraphics[width=0.5\textwidth]{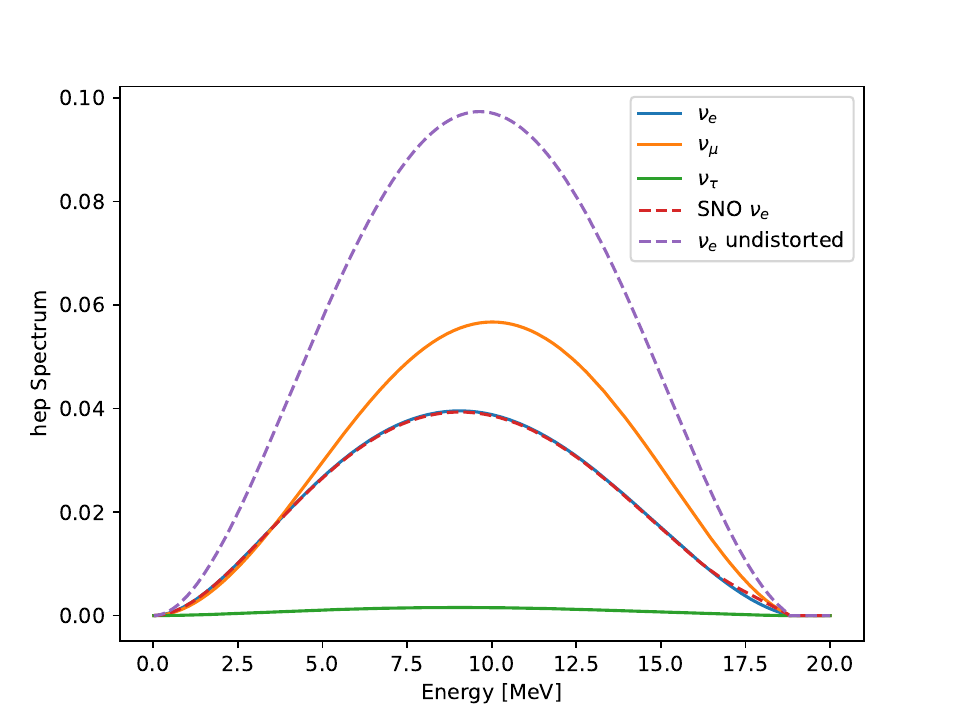}
\caption{Neutrino spectrum for the $^8B$ (left) and $hep$ (right) neutrino fractions at the surface of the Sun. Solid lines ares are the \peanuts predictions for the distorted spectrum (with oscillations). The dotted lines are the undistorted spectrum (no oscillations). Dashed lines are computing distorting the spectrum with the survival probability from the digitised curves for the SNO experiment, from Fig. 6.3 in~\cite{FiuzadeBarros:2011qna}.}
\label{fig:SNO_8B_hep_comparison_spectrum}
\end{figure}

In addition to comparing the survival probability, we can compare the distorted (i.e. including oscillations) energy spectrum of neutrinos at the surface of the Sun. Figure~\ref{fig:SNO_8B_hep_comparison_spectrum} shows the effect of the oscillation distortion on the electron neutrino spectrum, as well as the comparison with the spectrum distorted by the SNO survival probability. For this comparison we have used the spectra in \cite{Ortiz:2000nf} for the $^8B$ fraction and \cite{Bahcall:1997eg} for the $hep$ fraction.

\begin{figure}[ht]
 \includegraphics[width=0.5\textwidth]{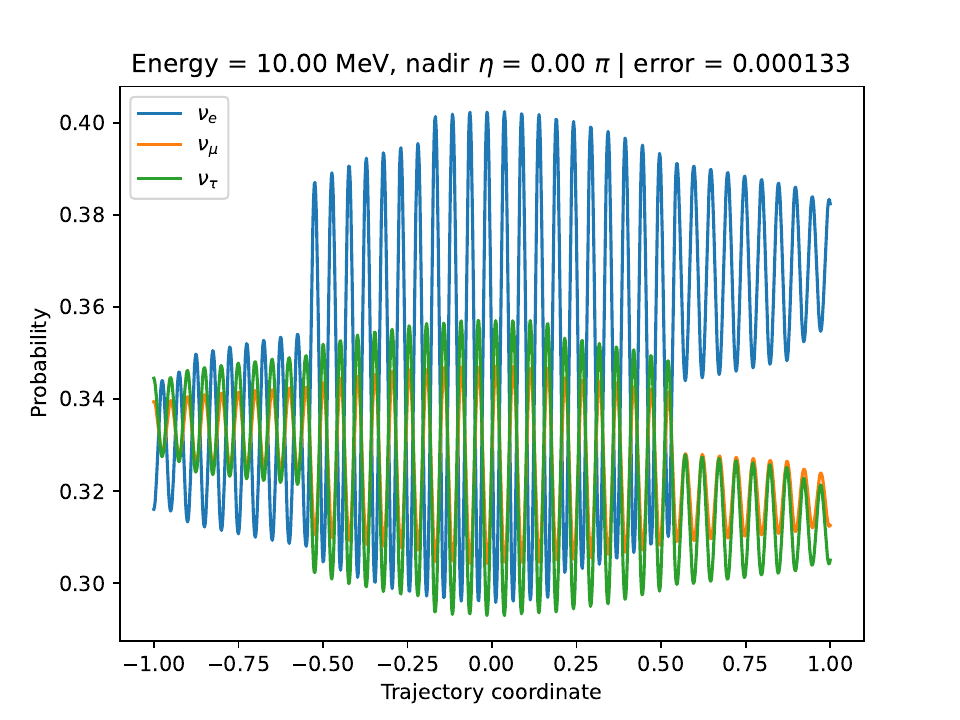}
 \includegraphics[width=0.5\textwidth]{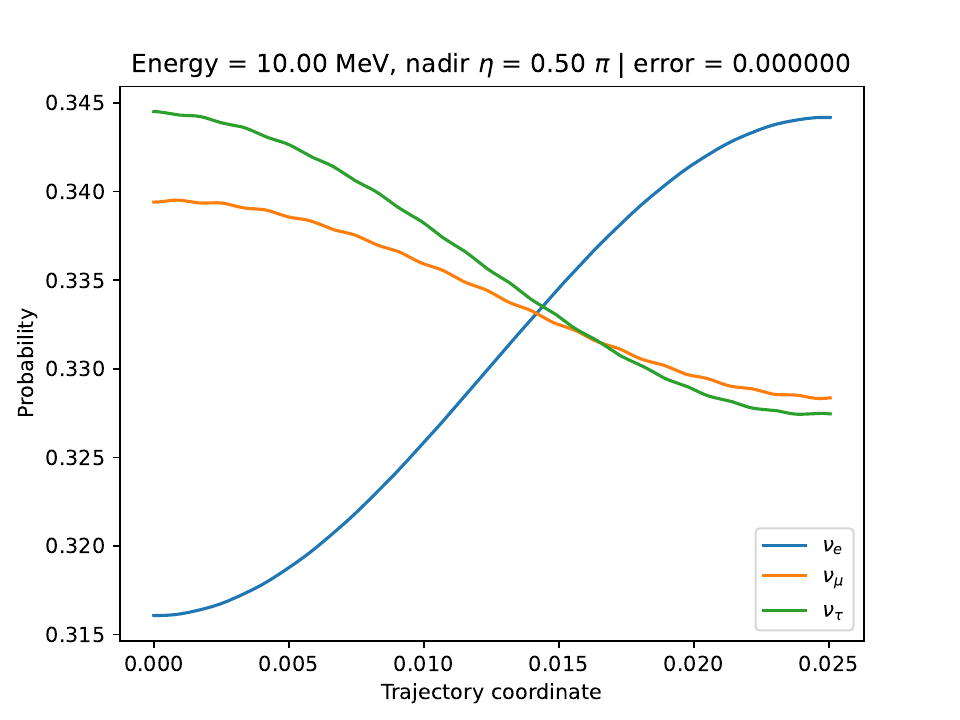}
\caption{Oscillations of neutrino flavour eigenstates for $\eta = 0$ (left) and $\eta = \pi/2$ (right). The error quoted is the difference between the analytical and numerical results.}
\label{fig:earth_regeneration}
\end{figure}

\begin{figure}[ht]
 \includegraphics[width=0.5\textwidth]{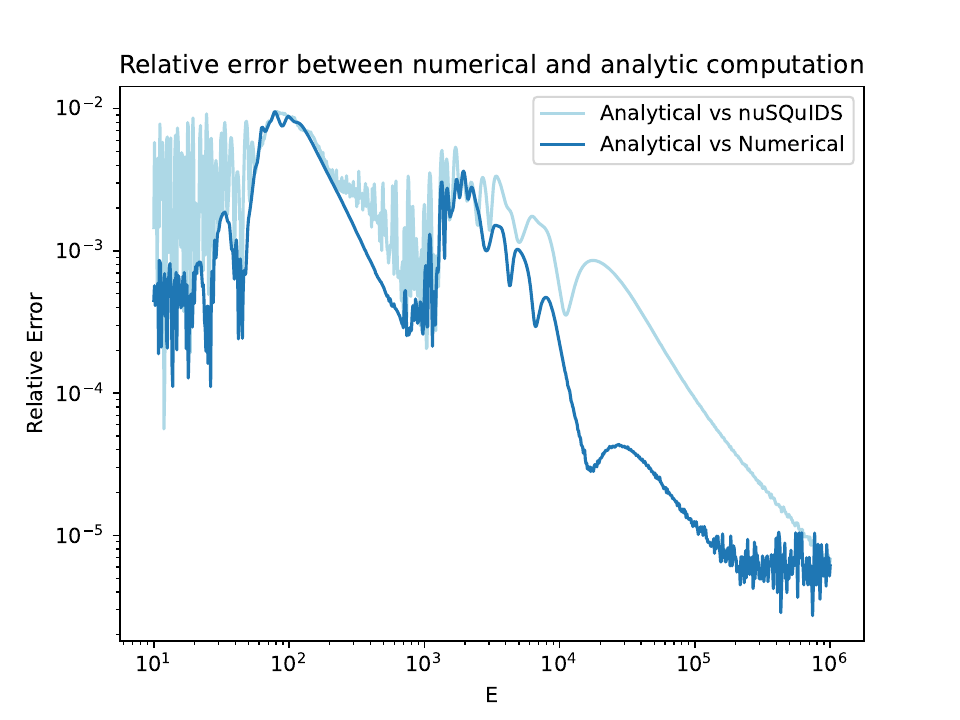}
 \includegraphics[width=0.5\textwidth]{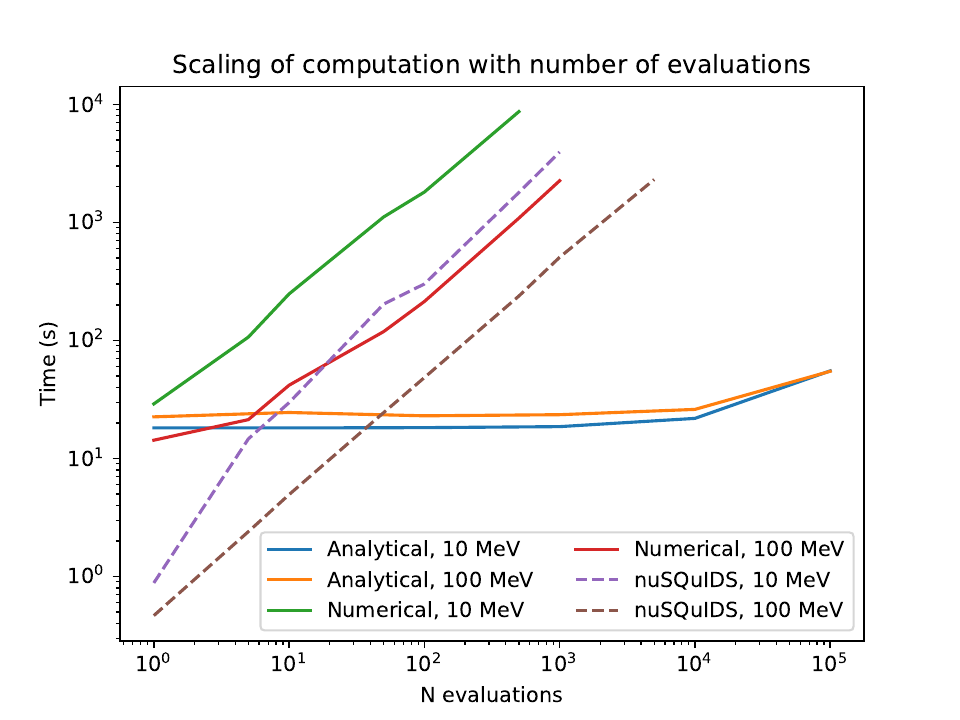}
\caption{Relative error between the numerical and analytical computations of the probability after Earth regeneration as a function of the energy for a chosen value of $\eta = 0$ (left). Speed comparison of the analytical and numerical computations by \peanuts and the computations by \textsf{nuSQuIDS} for two values of the neutrino energy (right).}
\label{fig:error}
\end{figure}

In addition to validating the results against those of the SNO experiment, we have also validated our analytical approximation from Section \ref{sec:perturb} against the full numerical calculation described at the beginning of Section \ref{sec:propag}. Figure \ref{fig:earth_regeneration} shows the oscillation pattern of the neutrino flavour eigenstates for a neutrino crossing all Earth shells, $\eta=0$ (left), and for a neutrino coming from the horizon, $\eta=\pi/2$ (right), starting from a pure mass neutrino eigenstate $\nu_\alpha = U_{\alpha 2}$. As expected, only in the first case the oscillations are significant, since the neutrinos traverse the whole Earth before reaching the detector, whereas in the second case the neutrino path only crosses part of Earth's crust between the surface and the detector, and thus only minimal oscillation occurs. Since only the numerical computation can provide the full trajectory, the oscillations shown in Fig.~\ref{fig:earth_regeneration} correspond exclusively to the numerical mode. However, for the same choice of parameters we have also used the analytical computation for the final probabilities, and the error reported in the figures precisely corresponds to the relative difference between the numerical and analytical models\footnote{Given $(c_e, c_\mu, c_\tau)$ the complex coefficients defining the final (evolved) state (cf. eq.~\ref{eq:general_flavour_state}), the relative error is defined here as the norm of the difference between the numerical and analytical values of it, divided by the norm of their sum. This ensures that errors in both real and imaginary parts of the solution are correctly taken into account.}. The difference is almost negligible, of the order of $10^{-4}$ for the night period (left), and effectively zero during the day period. The left panel in Figure \ref{fig:error} confirms this by showing the relative error as a function of the energy for the worse case of the ones above (i.e. during the night) for $\eta=0$, where it can clearly be noticed that the error is small for all values of the energy, and only approaches $\sim 10^{-2}$ at the worst for $E_\nu \sim 10^2$ MeV. This comparison serves as a validation that the approximated analytical solution is a very good approximation and, since it is much faster, can be used in place of the full numerical evaluation\footnote{Notice that, at energies above the TeV scale, neutrino inelastic scattering becomes relevant for Earth-type densities~\cite{IceCube:2017roe}. \peanuts does not currently include such effects. Fig.~\ref{fig:error} only compares the coherent forward scattering computed numerically and analytically.}. For completeness, for the production of Fig.~\ref{fig:error}, the unitarity of the evolution matrix was confirmed, and no unphysical solutions were found. Figure \ref{fig:error} also shows a comparison with the results from the state-of-the-art tool for neutrino oscillations \textsf{nuSQuIDS}~\cite{Arguelles:2021twb}, where it can be seen that the agreement is also good with a relative error consistently below $10^{-2}$. Therefore, the massive increase in computational speed provided by \peanuts certainly justifies the small loss in precision.\footnote{We choose not to compare our results with other neutrino tools, such as \textsf{GLoBES}\cite{Huber:2004ka, Huber:2007ji}, \textsf{Prob3++}\cite{Prob3pp} or \textsf{nuCRAFT}~\cite{Wallraff:2014qka}, as their main purpose is not solar neutrinos, but rather long baseline experiments, for the first two, and atmospheric neutrinos, for the latter, whereas \textsf{nuSQuIDS} is a general purpose tool.}

To quantify this speed increase gained with the analytical implementation, we show in the right panel of Figure \ref{fig:error} the computational time of the numerical and analytical computations as a function of number of evaluations, for two values of the neutrino energy $E_\nu = 10$ MeV and $E_\nu = 100$ MeV. For small number of evaluations the computational time is very similar between the analytical and numerical methods, but for a number of evaluations $N \gtrsim 10$, the computational time for the numerical method increases drastically and soon becomes computationally unfeasible. In contrast the total CPU time of the analytical method remains constant for increasing number of evaluations, and it is mostly dominated by the overhead of the initialisation step. Only for $N \gtrsim 10^5$ the computation time starts to increase noticeably, but still remains manageable up to large $N$. To emphasise further the increase of speed with the analytical implementation, the relative error shown on the left-hand panel of Figure \ref{fig:error} required over $9.2 \times 10^3$ seconds of CPU time to perform the numerical computations for all energies on a 2,3 GHz Intel Core i7 quad-core, compared to 5 seconds with the analytical method. Lastly, the right panel of Figure \ref{fig:error} also shows a comparison with \textsf{nuSQuIDS}. For low number of evaluations, \textsf{nuSQuIDS} clearly outperforms \peanuts, by virtue of being written in \textsf{C++}, but for large number of evaluations, as early as $N\gtrsim 5$ for low energies and $N\gtrsim 50$ for high energies, the analytical implementation of \peanuts becomes significantly faster and, again, is the only feasible option for these large number of evaluations. It is worth noting that this comparison is done in the worst case scenario, where $\eta=0$ and the neutrino path crosses almost the whole Earth. In less extreme scenarios, the computational speed of the numerical implementation of \peanuts and the computations by \textsf{nuSQuIDS} are somewhat faster than shown.

\begin{figure}[ht]
\centering
 \includegraphics[width=0.8\textwidth]{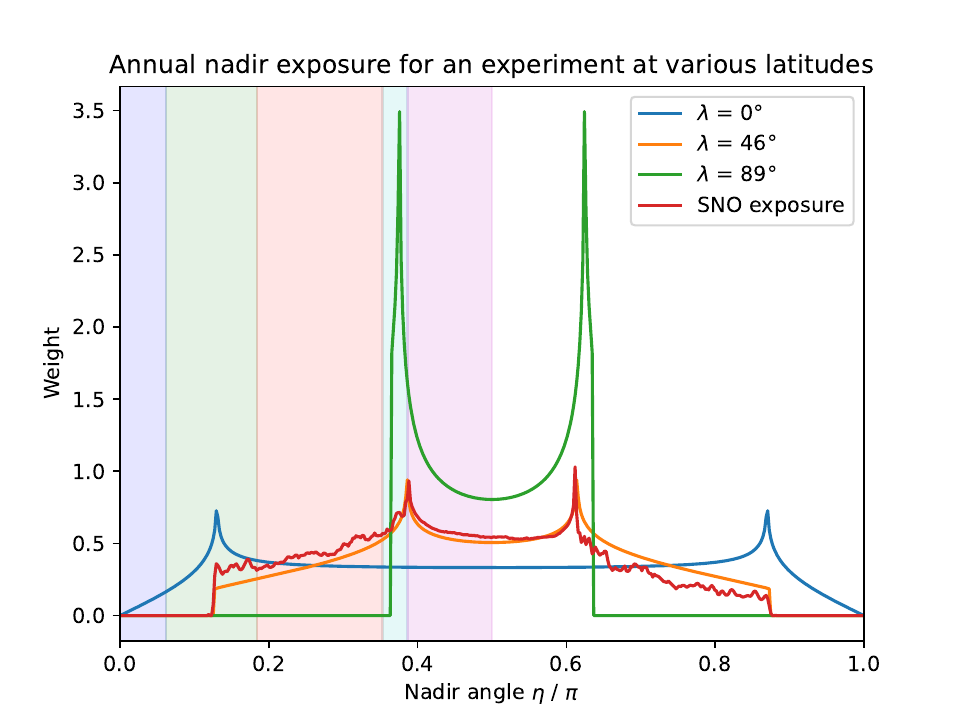}
\caption{Exposure of the SNO experiment in nadir angle $\eta$ (red), compared to the ideal exposure for hypothetical experiments at various latitudes. Coloured bands are as in Figure \ref{fig:eta_weights_ideal}.}
\label{fig:eta_weights}
\end{figure}

Section \ref{sec:time_integration} showed how to compute the ideal exposure time for a hypothetical experiment. For specific experiments, however, the exposure is often provided tabulated in bins of either the nadir angle $\eta$, the zenith angle $\theta$ or $\cos\theta$. In the case of the SNO experiment, the exposure is provided in bins of $\cos\theta$, hence we convert it into bins of $\eta$ in order to match fit our computations of the probability. We then show in Figure \ref{fig:eta_weights} the exposure of the SNO experiment compared to the ideal case. As SNO is located at a latitude of $46.475^{\circ}$, we can see that it matches very well the ideal exposure at $46^{\circ}$, with a slight under-exposure during the day and a slight over-exposure, during the night, which is consistent with the livetime of the SNO experiment.\footnote{This pattern is due to most of the maintenance and calibration operations taking place during daytime~\cite{FiuzadeBarros:2011qna}.} 

\begin{figure}[ht]
\centering
 \includegraphics[width=0.8\textwidth]{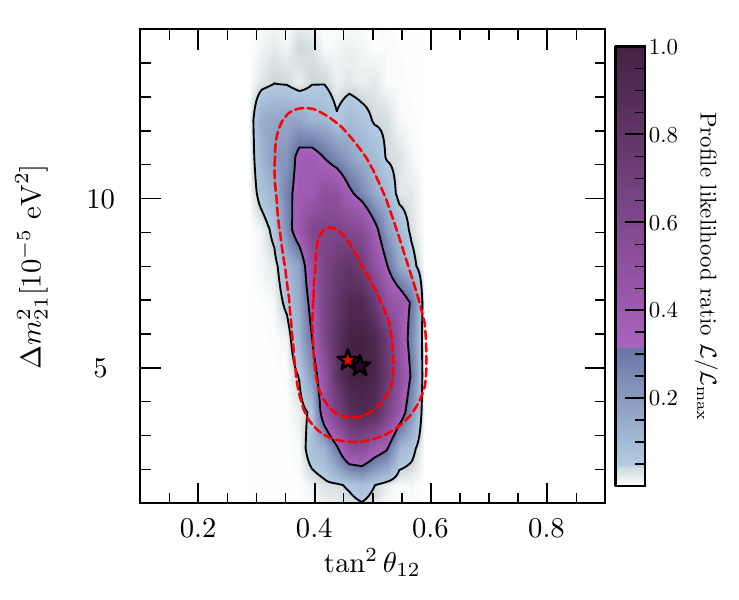}
\caption{Profile likelihood ratio for the results of a statistical fit of the oscillation parameters $\theta_{12}$ and $\Delta m_{21}^2$ using  \peanuts (purple-blue) compared to the results reported by the Phase I of the SNO experiment (red). The fit was performed using \textsf{GAMBIT}~\cite{GAMBIT:2017yxo} and the figure was generated with \textsf{pippi}~\cite{Scott:2012qh}.}
\label{fig:SNO_3_4_like2D}
\end{figure}

Finally, in order to match the computations of \peanuts with that of the SNO experiment, we reproduce the statistical fit of the oscillation parameters $\theta_{12}$ and $\Delta m_{21}^2$ performed by the SNO experiment. Figure \ref{fig:SNO_3_4_like2D} shows the results of the fit as a profile likelihood ratio. The dark purple star corresponds to the best fit point as found using \peanuts, while the red star is that reported by the SNO collaboration. Similarly the purple and blue shared contours correspond to the 68\% and 95\% confidence intervals around the best fit point for our results, whereas the red dashed contours are the same from the SNO results. Note that this comparison, both our results and those from the SNO experiment, corresponds only to Phase I of the experiment~\cite{SNO:2006odc}, and for normal ordering of neutrino masses\footnote{We have repeated the fit for inverted ordering and found the results to be almost identical to that of normal ordering, which is expected given that the oscillation of solar neutrinos is largely independent of $\Delta m^2_{3l}$.}. Details about this comparison will appear in a global fit by the \textsf{GAMBIT} Neutrino working group~\cite{GAMBITNus}. The results show a decent match with those reported by the SNO collaboration, with the best fit points laying very close to each other. The shape and reach of the contours is larger in our study, which can be attributed to a slight difference on the treatment of systematic uncertainties. It is crucial here to emphasize that a parameter scan of this magnitude is only feasible with the analytical implementation of \peanuts, due to the large number of evaluations required. For reference, the scan sampled around 380k parameter points, each of which performed, on average, around 5k evaluations of the probability. 

\section{Conclusions and Outlook}
\label{sec:conclusions}

We have presented in this paper \peanuts, a fast and flexible software to automatically compute the energy spectra of solar active neutrinos, for arbitrary solar models and custom Earth density profiles. \peanuts assumes adiabatic propagation of neutrinos within the Sun, and provides analytic computation for the coherent evolution of active neutrinos while crossing the Earth, thus completely avoiding any time-consuming numerical integration. This, together with extensive use of pre-compiled and just-in-time compilation optimisations, makes the software extremely fast and optimised for large-scale parameter scans.

\peanuts provides algorithms to automatically perform the full chain of computations to simulate a solar neutrino experiment, as well as easy individual access to the modules and functions for specific computations. These include, for instance: mixing parameters in matter, incoherent flux at Sun surface, evolved neutrino state after Earth crossing, Earth density profile for given nadir angle, evolutor operator for given neutrino energy and nadir angle, nadir exposure for an experiment between two arbitrary days of the year, integrated probability of oscillation over a finite observation time. In the present version of \peanuts we focused on providing automatisation for solar neutrinos, but the modularity of \peanuts also allows a user to employ, for example, the function \py{Pearth_analytical} to compute the evolution of an atmospheric neutrino, or simulate the evolved spectra for hypothetical neutrinos produced in solar flares.

\peanuts can be run in \emph{simple} mode, for quick computations directly from the command line, as well as in \emph{expert} mode, in which case the user provides a set of comprehensive instructions in the form of a \yaml file. The \emph{expert} mode is thus ideal for scripting, as it exploits all the possible functionalities of \peanuts, and to perform simple explorations of the parameter space, as it natively allows the scanning of various input parameters on a grid. 

We extensively validated \peanuts against the results of the SNO experiment, and provide ready-to-run scripts to reproduce our results. We observed excellent agreement for the survival probability at Sun surface, distorted neutrino spectra for the ${}^8B$ and $hep$ channels, annual nadir exposure and ability to reproduce the profile likelihood of the Phase I of the experiment for the parameters $\theta_{12}$ and $\Delta m^2_{21}$. We also validated our analytic computation of the evolved neutrino state with a numerical solution, and find excellent agreement on a wide range of energies. We performed a quick estimation of the CPU time scaling of the analytical and numerical computations with the number of evaluations, and conclusively found our analytical implementation to significantly outperform the numerical one for as few as $\mathcal{O}(10)$ evaluations. We also compared our results with those of \textsf{nuSQuIDS} and found good agreement as well. As with the numerical implementation, we found that our analytical solution vastly outperforms \textsf{nuSQuIDS} in terms of the computational time. \peanuts thus presents a very fast alternative to existing tools without appreciable loss of precision. Therefore, due to its speed and user-friendly interface, we argue that \peanuts is a very capable and useful tool for the computation of the propagation of neutrinos in the Sun and through Earth and a crucial addition to the software toolbox of the neutrino physics community. 

Concerning the limitations of the software, the current \peanuts version assumes adiabatic evolution within the Sun, which provides an excellent approximation for solar neutrino energies given the currently allowed range of oscillation parameters; the user is thus encouraged to check the validity of this regime if working in non-standard scenarios. We also stress that the code assumes coherent forward scattering for the propagation of neutrinos in matter: for Earth-type densities, inelastic scattering becomes important at energies above the TeV scale, and such effects must thus be taken into account. We plan to incorporate a more general routine for the arbitrary evolution of neutrinos in the Sun in a future version of \peanuts. For inelastic scattering, other softwares are currently available, cf. e.g.~\cite{Gazizov:2004va,Vincent:2017svp,Garcia:2020jwr,Safa:2021ghs,Arguelles:2021twb,Garg:2022ugd}. Though \peanuts v1.0 does not yet provide all the functionalities that other software tools have, it vastly outperforms them in speed, which can be a critical factor in e.g. global studies, where the number of evaluations is very large.

Further improvements that will extend the scope of \peanuts are under investigation, most notably the implementation of algorithms for the automatic computation of the atmospheric neutrino flux at a given location, as well as the simulation of accelerator neutrinos beams. 

\section*{Acknowledgements}

We would like to thank Aaron Vincent for helpful comments about this work, Roberto Ruiz de Austri for testing and suggestions, and the rest of the GAMBIT community for useful discussions. We also thank Will Handley for access to HPC resources, as for this work we used the Cambridge Service for Data Driven Discovery(CSD3), part of which is operated by the University of Cambridge Research Computing on behalf of the STFC DiRAC HPC Facility (www.dirac.ac.uk). The DiRAC component of CSD3 was funded by BEIS capital funding via STFC capital grants ST/P002307/1 and ST/R002452/1 and STFC operations grant ST/R00689X/1. DiRAC is part of the National e-Infrastructure. ML is funded by the European Union under the Horizon Europe's Marie Sklodowska-Curie project 101068791 — NuBridge. ML acknowledges financial support from the Alexander von Humboldt Foundation during the early stages of the work, and thanks the UCLouvain Centre for Cosmology, Particle Physics and Phenomenology (CP3) for hosting him during the final stage. TEG is funded by the Deutsche Forschungsgemeinschaft (DFG) through the Emmy Noether Grant No. KA 4662/1-1. 

\appendix

\section{Installation guide with the \texttt{conda} environment}
\label{app:installation}
In this Appendix we provide a short guide to install and run \peanuts from scratch in a system-agnostic way, using the \texttt{conda}~\cite{anaconda} and \texttt{pip}~\cite{pypi} package managers.

First of all the user shall create a separate working environment, to avoid conflicts with existing packages:
\begin{lstpy}
conda create -n peanuts python numpy numba scipy pandas mpmath pyyaml matplotlib gitpython
conda activate peanuts
pip install pyinterval pyslha
\end{lstpy}
The first line will create a new \texttt{conda} environment named \texttt{peanuts}, simultaneously installing in it the listed packages ensuring mutual compatibility in their versions. The second line activates this environment, while the third one installs the additional required packages that are not currently available via \texttt{conda}, using instead the \texttt{pip} package manager.

The user can now download \peanuts and move to the folder containing its \texttt{Python} scripts:
\begin{lstpy}
git clone https://github.com/michelelucente/PEANUTS
cd PEANUTS 
\end{lstpy}

At this point \peanuts should be installed and ready to run. The user can test it on the provided example input files, for instance:
\begin{lstpy}
python run_peanuts.py -f examples/solar_earth_test.yaml
\end{lstpy}

The above-described procedure installs the latest version of the packages that are mutually compatible at a given time. It has been successfully tested at time of publication, and the resulting environment is provided in the \texttt{PEANUTS} folder as \texttt{peanuts\_env.yml}. The user can reproduce this installation by running
\begin{lstpy}
conda env create -f peanuts_env.yml
\end{lstpy}
This command will take care of installing both the \texttt{pip} and \texttt{conda} packages with versions as defined in \texttt{peanuts\_env.yml}.

\bibliographystyle{JHEP}
\bibliography{bibliography.bib}

\providecommand{\href}[2]{#2}\begingroup\raggedright\begin{thebibliography}{10}

\bibitem{Bahcall:1989ks}
J.N.~Bahcall, \emph{{NEUTRINO ASTROPHYSICS}}, Cambridge University Press
  (1989).

\bibitem{Bahcall:2004pz}
J.N.~Bahcall, A.M.~Serenelli and S.~Basu, \emph{{New solar opacities,
  abundances, helioseismology, and neutrino fluxes}},
  \href{https://doi.org/10.1086/428929}{\emph{Astrophys. J. Lett.} {\bfseries
  621} (2005) L85} [\href{https://arxiv.org/abs/astro-ph/0412440}{{\ttfamily
  astro-ph/0412440}}].

\bibitem{Bahcall:1981br}
J.N.~Bahcall and R.~Davis, Jr., \emph{{AN ACCOUNT OF THE DEVELOPMENT OF THE
  SOLAR NEUTRINO PROBLEM}}, .

\bibitem{Davis:1968cp}
R.~Davis, Jr., D.S.~Harmer and K.C.~Hoffman, \emph{{Search for neutrinos from
  the sun}}, \href{https://doi.org/10.1103/PhysRevLett.20.1205}{\emph{Phys.
  Rev. Lett.} {\bfseries 20} (1968) 1205}.

\bibitem{Bahcall:1963ohf}
J.N.~Bahcall, W.A.~Fowler, I.~Iben, Jr. and R.L.~Sears, \emph{{Solar neutrino
  flux}}, \href{https://doi.org/10.1086/147513}{\emph{Astrophys. J.} {\bfseries
  137} (1963) 344}.

\bibitem{Sears:1964zz}
R.L.~Sears, \emph{{Helium Content and Neutrino Fluxes in Solar Models}},
  \href{https://doi.org/10.1086/147942}{\emph{Astrophys. J.} {\bfseries 140}
  (1964) 477}.

\bibitem{Pochoda:1964ana}
P.~Pochoda and H.~Reeves, \emph{{A revised solar model with a solar neutrino
  spectrum}}, \href{https://doi.org/10.1016/0032-0633(64)90116-3}{\emph{Planet.
  Space Sci.} {\bfseries 12} (1964) 119}.

\bibitem{Bahcall:1964gx}
J.N.~Bahcall, \emph{{Solar neutrinos. I: Theoretical}},
  \href{https://doi.org/10.1103/PhysRevLett.12.300}{\emph{Phys. Rev. Lett.}
  {\bfseries 12} (1964) 300}.

\bibitem{ez06000k}
D.~Ezer and A.G.W.~Cameron, \emph{A study of solar evolution},
  \href{https://doi.org/10.1139/p65-140}{\emph{Can. J. Phys.} {\bfseries 43}
  (1965) 1497}.

\bibitem{ez07000g}
D.~Ezer and A.G.W.~Cameron, \emph{Solar evolution with varying g},
  \href{https://doi.org/10.1139/p66-050}{\emph{Can. J. Phys.} {\bfseries 44}
  (1966) 593}.

\bibitem{1967ApJ...150..723B}
J.N.~{Bahcall}, M.~{Cooper} and P.~{Demarque}, \emph{{Dependence of the
  \^8\{B\} Solar Neutrino Flux on Heavy Element Composition}},
  \href{https://doi.org/10.1086/149373}{\emph{ApJ} {\bfseries 150} (1967) 723}.

\bibitem{1967ApJ...150..725S}
G.~{Shaviv}, J.N.~{Bahcall} and W.A.~{Fowler}, \emph{{Dependence of the
  \^8\{B\} Solar Neutrino Flux on the Rate of the Reaction
  \^3He(\^3He,2p)$^{4}$He}}, \href{https://doi.org/10.1086/149374}{\emph{ApJ}
  {\bfseries 150} (1967) 725}.

\bibitem{Bahcall:1968jvj}
J.N.~Bahcall, N.A.~Bahcall, W.A.~Fowler and G.~Shaviv, \emph{{Solar neutrinos
  and low-energy nuclear cross sections}},
  \href{https://doi.org/10.1016/0370-2693(68)90610-2}{\emph{Phys. Lett. B}
  {\bfseries 26} (1968) 359}.

\bibitem{SNO:2001kpb}
{\scshape SNO} collaboration, \emph{{Measurement of the rate of $\nu_e+d \to
  p+p+e^-$ interactions produced by $^8$B solar neutrinos at the Sudbury
  Neutrino Observatory}},
  \href{https://doi.org/10.1103/PhysRevLett.87.071301}{\emph{Phys. Rev. Lett.}
  {\bfseries 87} (2001) 071301}
  [\href{https://arxiv.org/abs/nucl-ex/0106015}{{\ttfamily nucl-ex/0106015}}].

\bibitem{Maki:1962mu}
Z.~Maki, M.~Nakagawa and S.~Sakata, \emph{{Remarks on the unified model of
  elementary particles}}, \href{https://doi.org/10.1143/PTP.28.870}{\emph{Prog.
  Theor. Phys.} {\bfseries 28} (1962) 870}.

\bibitem{Pontecorvo:1967fh}
B.~Pontecorvo, \emph{{Neutrino Experiments and the Problem of Conservation of
  Leptonic Charge}}, {\emph{Zh. Eksp. Teor. Fiz.} {\bfseries 53} (1967) 1717}.

\bibitem{Gribov:1968kq}
V.N.~Gribov and B.~Pontecorvo, \emph{{Neutrino astronomy and lepton charge}},
  \href{https://doi.org/10.1016/0370-2693(69)90525-5}{\emph{Phys. Lett. B}
  {\bfseries 28} (1969) 493}.

\bibitem{Esteban:2020cvm}
I.~Esteban, M.C.~Gonzalez-Garcia, M.~Maltoni, T.~Schwetz and A.~Zhou,
  \emph{{The fate of hints: updated global analysis of three-flavor neutrino
  oscillations}}, \href{https://doi.org/10.1007/JHEP09(2020)178}{\emph{JHEP}
  {\bfseries 09} (2020) 178}
  [\href{https://arxiv.org/abs/2007.14792}{{\ttfamily 2007.14792}}].

\bibitem{deSalas:2020pgw}
P.F.~de~Salas, D.V.~Forero, S.~Gariazzo, P.~Mart{\'\i}nez-Mirav{\'e}, O.~Mena,
  C.A.~Ternes et~al., \emph{{2020 global reassessment of the neutrino
  oscillation picture}},
  \href{https://doi.org/10.1007/JHEP02(2021)071}{\emph{JHEP} {\bfseries 02}
  (2021) 071} [\href{https://arxiv.org/abs/2006.11237}{{\ttfamily
  2006.11237}}].

\bibitem{Capozzi:2017ipn}
F.~Capozzi, E.~Di~Valentino, E.~Lisi, A.~Marrone, A.~Melchiorri and A.~Palazzo,
  \emph{{Global constraints on absolute neutrino masses and their ordering}},
  \href{https://doi.org/10.1103/PhysRevD.95.096014}{\emph{Phys. Rev. D}
  {\bfseries 95} (2017) 096014}
  [\href{https://arxiv.org/abs/2003.08511}{{\ttfamily 2003.08511}}].

\bibitem{BOREXINO:2020aww}
{\scshape BOREXINO} collaboration, \emph{{Experimental evidence of neutrinos
  produced in the CNO fusion cycle in the Sun}},
  \href{https://doi.org/10.1038/s41586-020-2934-0}{\emph{Nature} {\bfseries
  587} (2020) 577} [\href{https://arxiv.org/abs/2006.15115}{{\ttfamily
  2006.15115}}].

\bibitem{Borexino:2017rsf}
{\scshape Borexino} collaboration, \emph{{First Simultaneous Precision
  Spectroscopy of $pp$, $^7$Be, and $pep$ Solar Neutrinos with Borexino
  Phase-II}}, \href{https://doi.org/10.1103/PhysRevD.100.082004}{\emph{Phys.
  Rev. D} {\bfseries 100} (2019) 082004}
  [\href{https://arxiv.org/abs/1707.09279}{{\ttfamily 1707.09279}}].

\bibitem{BOREXINO:2018ohr}
{\scshape BOREXINO} collaboration, \emph{{Comprehensive measurement of
  $pp$-chain solar neutrinos}},
  \href{https://doi.org/10.1038/s41586-018-0624-y}{\emph{Nature} {\bfseries
  562} (2018) 505}.

\bibitem{Yano:2020aap}
{\scshape Hyper-Kamiokande Proto} collaboration, \emph{{Solar neutrino physics
  at Hyper-Kamiokande}}, \href{https://doi.org/10.22323/1.358.1037}{\emph{PoS}
  {\bfseries ICRC2019} (2020) 1037}.

\bibitem{IceCube:2021jwt}
{\scshape IceCube} collaboration, \emph{{Search for GeV neutrino emission
  during intense gamma-ray solar flares with the IceCube Neutrino
  Observatory}}, \href{https://doi.org/10.1103/PhysRevD.103.102001}{\emph{Phys.
  Rev. D} {\bfseries 103} (2021) 102001}
  [\href{https://arxiv.org/abs/2101.00610}{{\ttfamily 2101.00610}}].

\bibitem{Super-Kamiokande:2022yrk}
{\scshape Super-Kamiokande} collaboration, \emph{{Searching for neutrinos from
  solar flares across solar cycles 23 and 24 with the Super-Kamiokande
  detector}},  \href{https://arxiv.org/abs/2210.12948}{{\ttfamily 2210.12948}}.

\bibitem{IceCube:2021koo}
{\scshape IceCube} collaboration, \emph{{Recent Progress in Solar Atmospheric
  Neutrino Searches with IceCube}},
  \href{https://doi.org/10.22323/1.395.1174}{\emph{PoS} {\bfseries ICRC2021}
  (2021) 1174} [\href{https://arxiv.org/abs/2107.13696}{{\ttfamily
  2107.13696}}].

\bibitem{In:2017kcf}
{\scshape IceCube} collaboration, \emph{{Latest results and sensitivities for
  solar dark matter searches with IceCube}},
  \href{https://doi.org/10.22323/1.301.0912}{\emph{PoS} {\bfseries ICRC2017}
  (2018) 912}.

\bibitem{Super-Kamiokande:2002ujc}
{\scshape Super-Kamiokande} collaboration, \emph{{Determination of solar
  neutrino oscillation parameters using 1496 days of Super-Kamiokande I data}},
  \href{https://doi.org/10.1016/S0370-2693(02)02090-7}{\emph{Phys. Lett. B}
  {\bfseries 539} (2002) 179}
  [\href{https://arxiv.org/abs/hep-ex/0205075}{{\ttfamily hep-ex/0205075}}].

\bibitem{Bahcall:2003ce}
J.N.~Bahcall and C.~Pena-Garay, \emph{{Global analyses as a road map to solar
  neutrino fluxes and oscillation parameters}},
  \href{https://doi.org/10.1088/1126-6708/2003/11/004}{\emph{JHEP} {\bfseries
  11} (2003) 004} [\href{https://arxiv.org/abs/hep-ph/0305159}{{\ttfamily
  hep-ph/0305159}}].

\bibitem{Super-Kamiokande:2022lyl}
{\scshape Super-Kamiokande} collaboration, \emph{{Testing Non-Standard
  Interactions Between Solar Neutrinos and Quarks with Super-Kamiokande}},
  \href{https://arxiv.org/abs/2203.11772}{{\ttfamily 2203.11772}}.

\bibitem{Super-Kamiokande:2020frs}
{\scshape Super-Kamiokande} collaboration, \emph{{Search for solar electron
  anti-neutrinos due to spin-flavor precession in the Sun with
  Super-Kamiokande-IV}},
  \href{https://doi.org/10.1016/j.astropartphys.2022.102702}{\emph{Astropart.
  Phys.} {\bfseries 139} (2022) 102702}
  [\href{https://arxiv.org/abs/2012.03807}{{\ttfamily 2012.03807}}].

\bibitem{Wolfenstein:1977ue}
L.~Wolfenstein, \emph{{Neutrino Oscillations in Matter}},
  \href{https://doi.org/10.1103/PhysRevD.17.2369}{\emph{Phys. Rev. D}
  {\bfseries 17} (1978) 2369}.

\bibitem{Mikheyev:1985zog}
S.P.~Mikheyev and A.Y.~Smirnov, \emph{{Resonance Amplification of Oscillations
  in Matter and Spectroscopy of Solar Neutrinos}}, {\emph{Sov. J. Nucl. Phys.}
  {\bfseries 42} (1985) 913}.

\bibitem{FiuzadeBarros:2011qna}
N.F.~Fi\'uza~de Barros, \emph{{Precision Measurement of Neutrino Oscillation
  Parameters: Combined Three-phase Results of the Sudbury Neutrino
  Observatory}}, Ph.D. thesis, Lisbon U., 2011.

\bibitem{SNO:2006odc}
{\scshape SNO} collaboration, \emph{{Determination of the $\nu_e$ and total
  $^8$B solar neutrino fluxes with the Sudbury neutrino observatory phase I
  data set}}, \href{https://doi.org/10.1103/PhysRevC.75.045502}{\emph{Phys.
  Rev. C} {\bfseries 75} (2007) 045502}
  [\href{https://arxiv.org/abs/nucl-ex/0610020}{{\ttfamily nucl-ex/0610020}}].

\bibitem{Bahcall:2000nu}
J.N.~Bahcall, M.H.~Pinsonneault and S.~Basu, \emph{{Solar models: Current epoch
  and time dependences, neutrinos, and helioseismological properties}},
  \href{https://doi.org/10.1086/321493}{\emph{Astrophys. J.} {\bfseries 555}
  (2001) 990} [\href{https://arxiv.org/abs/astro-ph/0010346}{{\ttfamily
  astro-ph/0010346}}].

\bibitem{Mikheev:1987wa}
S.P.~Mikheev and A.Y.~Smirnov, \emph{{NEUTRINO OSCILLATIONS IN MATTER}},  in
  \emph{{International Symposium on Weak and Electromagnetic Interactions in
  Nuclei}}, pp.~405--415, 1987.

\bibitem{Lisi:1997yc}
E.~Lisi and D.~Montanino, \emph{{Earth regeneration effect in solar neutrino
  oscillations: An Analytic approach}},
  \href{https://doi.org/10.1103/PhysRevD.56.1792}{\emph{Phys. Rev. D}
  {\bfseries 56} (1997) 1792}
  [\href{https://arxiv.org/abs/hep-ph/9702343}{{\ttfamily hep-ph/9702343}}].

\bibitem{Bruss:1988fr}
D.~Bruss and L.M.~Sehgal, \emph{{DISTINGUISHING A COHERENT FROM AN INCOHERENT
  MIXTURE OF NEUTRINO FLAVORS}},
  \href{https://doi.org/10.1016/0370-2693(89)91144-1}{\emph{Phys. Lett. B}
  {\bfseries 216} (1989) 426}.

\bibitem{Mikheev:1986if}
S.P.~Mikheev and A.Y.~Smirnov, \emph{{Neutrino Oscillations in a Variable
  Density Medium and Neutrino Bursts Due to the Gravitational Collapse of
  Stars}}, {\emph{Sov. Phys. JETP} {\bfseries 64} (1986) 4}
  [\href{https://arxiv.org/abs/0706.0454}{{\ttfamily 0706.0454}}].

\bibitem{Denton:2016wmg}
P.B.~Denton, H.~Minakata and S.J.~Parke, \emph{{Compact Perturbative
  Expressions For Neutrino Oscillations in Matter}},
  \href{https://doi.org/10.1007/JHEP06(2016)051}{\emph{JHEP} {\bfseries 06}
  (2016) 051} [\href{https://arxiv.org/abs/1604.08167}{{\ttfamily
  1604.08167}}].

\bibitem{Denton:2018hal}
P.B.~Denton and S.J.~Parke, \emph{{Addendum to ``Compact perturbative
  expressions for neutrino oscillations in matter''}},
  \href{https://arxiv.org/abs/1801.06514}{{\ttfamily 1801.06514}}.

\bibitem{Ioannisian:2018qwl}
A.~Ioannisian and S.~Pokorski, \emph{{Three Neutrino Oscillations in Matter}},
  \href{https://doi.org/10.1016/j.physletb.2018.06.001}{\emph{Phys. Lett. B}
  {\bfseries 782} (2018) 641}
  [\href{https://arxiv.org/abs/1801.10488}{{\ttfamily 1801.10488}}].

\bibitem{Denton:2018mop}
P.B.~Denton, H.~Minakata and S.J.~Parke, \emph{{Comment on 1801.10488v3}}, .

\bibitem{Fantini:2018itu}
G.~Fantini, A.~Gallo~Rosso, F.~Vissani and V.~Zema, \emph{{Introduction to the
  Formalism of Neutrino Oscillations}},
  \href{https://doi.org/10.1142/9789813226098_0002}{\emph{Adv. Ser. Direct.
  High Energy Phys.} {\bfseries 28} (2018) 37}
  [\href{https://arxiv.org/abs/1802.05781}{{\ttfamily 1802.05781}}].

\bibitem{Kuo:1989qe}
T.-K.~Kuo and J.T.~Pantaleone, \emph{{Neutrino Oscillations in Matter}},
  \href{https://doi.org/10.1103/RevModPhys.61.937}{\emph{Rev. Mod. Phys.}
  {\bfseries 61} (1989) 937}.

\bibitem{Bethe:1991zq}
H.A.~Bethe and J.N.~Bahcall, \emph{{Solar neutrinos and the
  Mikheev-Smirnov-Wolfenstein theory}},
  \href{https://doi.org/10.1103/PhysRevD.44.2962}{\emph{Phys. Rev. D}
  {\bfseries 44} (1991) 2962}.

\bibitem{Harley:1992an}
D.~Harley, T.-K.~Kuo and J.T.~Pantaleone, \emph{{Solar neutrinos with three
  flavor mixings}}, \href{https://doi.org/10.1103/PhysRevD.47.4059}{\emph{Phys.
  Rev. D} {\bfseries 47} (1993) 4059}
  [\href{https://arxiv.org/abs/hep-ph/9208241}{{\ttfamily hep-ph/9208241}}].

\bibitem{Akhmedov:2004rq}
E.K.~Akhmedov, M.A.~Tortola and J.W.F.~Valle, \emph{{A Simple analytic three
  flavor description of the day night effect in the solar neutrino flux}},
  \href{https://doi.org/10.1088/1126-6708/2004/05/057}{\emph{JHEP} {\bfseries
  05} (2004) 057} [\href{https://arxiv.org/abs/hep-ph/0404083}{{\ttfamily
  hep-ph/0404083}}].

\bibitem{Vinyoles:2016djt}
N.~Vinyoles, A.M.~Serenelli, F.L.~Villante, S.~Basu, J.~Bergstr\"om,
  M.C.~Gonzalez-Garcia et~al., \emph{{A new Generation of Standard Solar
  Models}}, \href{https://doi.org/10.3847/1538-4357/835/2/202}{\emph{Astrophys.
  J.} {\bfseries 835} (2017) 202}
  [\href{https://arxiv.org/abs/1611.09867}{{\ttfamily 1611.09867}}].

\bibitem{Bahcall:1997eg}
J.N.~Bahcall, \emph{{Gallium solar neutrino experiments: Absorption
  cross-sections, neutrino spectra, and predicted event rates}},
  \href{https://doi.org/10.1103/PhysRevC.56.3391}{\emph{Phys. Rev. C}
  {\bfseries 56} (1997) 3391}
  [\href{https://arxiv.org/abs/hep-ph/9710491}{{\ttfamily hep-ph/9710491}}].

\bibitem{Ortiz:2000nf}
C.E.~Ortiz, A.~Garcia, R.A.~Waltz, M.~Bhattacharya and A.K.~Komives,
  \emph{{Shape of the B-8 alpha and neutrino spectra}},
  \href{https://doi.org/10.1103/PhysRevLett.85.2909}{\emph{Phys. Rev. Lett.}
  {\bfseries 85} (2000) 2909}
  [\href{https://arxiv.org/abs/nucl-ex/0003006}{{\ttfamily nucl-ex/0003006}}].

\bibitem{Bahcall:1987jc}
J.N.~Bahcall and R.K.~Ulrich, \emph{{Solar Models, Neutrino Experiments and
  Helioseismology}},
  \href{https://doi.org/10.1103/RevModPhys.60.297}{\emph{Rev. Mod. Phys.}
  {\bfseries 60} (1988) 297}.

\bibitem{Bahcall:1994cf}
J.N.~Bahcall, \emph{{The Be-7 solar neutrino line: A Reflection of the central
  temperature distribution of the sun}},
  \href{https://doi.org/10.1103/PhysRevD.49.3923}{\emph{Phys. Rev. D}
  {\bfseries 49} (1994) 3923}
  [\href{https://arxiv.org/abs/astro-ph/9401024}{{\ttfamily
  astro-ph/9401024}}].

\bibitem{Winter:2004kf}
W.T.~Winter, S.J.~Freedman, K.E.~Rehm and J.P.~Schiffer, \emph{{The B-8
  neutrino spectrum}},
  \href{https://doi.org/10.1103/PhysRevC.73.025503}{\emph{Phys. Rev. C}
  {\bfseries 73} (2006) 025503}
  [\href{https://arxiv.org/abs/nucl-ex/0406019}{{\ttfamily nucl-ex/0406019}}].

\bibitem{Dyson:1949bp}
F.J.~Dyson, \emph{{The Radiation theories of Tomonaga, Schwinger, and
  Feynman}}, \href{https://doi.org/10.1103/PhysRev.75.486}{\emph{Phys. Rev.}
  {\bfseries 75} (1949) 486}.

\bibitem{Ohlsson:1999xb}
T.~Ohlsson and H.~Snellman, \emph{{Three flavor neutrino oscillations in
  matter}}, \href{https://doi.org/10.1063/1.533270}{\emph{J. Math. Phys.}
  {\bfseries 41} (2000) 2768}
  [\href{https://arxiv.org/abs/hep-ph/9910546}{{\ttfamily hep-ph/9910546}}].

\bibitem{AbdusSalam:2020rdj}
S.S.~AbdusSalam et~al., \emph{{Simple and statistically sound recommendations
  for analysing physical theories}},
  \href{https://doi.org/10.1088/1361-6633/ac60ac}{\emph{Rept. Prog. Phys.}
  {\bfseries 85} (2022) 052201}
  [\href{https://arxiv.org/abs/2012.09874}{{\ttfamily 2012.09874}}].

\bibitem{GAMBITNus}
{\scshape GAMBIT Neutrino Workgroup} collaboration, \emph{Global analysis of
  neutrino oscillations with gambit}, {\emph{in preparation} (2023) }.

\bibitem{IceCube:2017roe}
{\scshape IceCube} collaboration, \emph{{Measurement of the multi-TeV neutrino
  cross section with IceCube using Earth absorption}},
  \href{https://doi.org/10.1038/nature24459}{\emph{Nature} {\bfseries 551}
  (2017) 596} [\href{https://arxiv.org/abs/1711.08119}{{\ttfamily
  1711.08119}}].

\bibitem{Arguelles:2021twb}
C.A.~Arg\"uelles, J.~Salvado and C.N.~Weaver, \emph{{nuSQuIDS: A toolbox for
  neutrino propagation}},
  \href{https://doi.org/10.1016/j.cpc.2022.108346}{\emph{Comput. Phys. Commun.}
  {\bfseries 277} (2022) 108346}
  [\href{https://arxiv.org/abs/2112.13804}{{\ttfamily 2112.13804}}].

\bibitem{Huber:2004ka}
P.~Huber, M.~Lindner and W.~Winter, \emph{{Simulation of long-baseline neutrino
  oscillation experiments with GLoBES (General Long Baseline Experiment
  Simulator)}}, \href{https://doi.org/10.1016/j.cpc.2005.01.003}{\emph{Comput.
  Phys. Commun.} {\bfseries 167} (2005) 195}
  [\href{https://arxiv.org/abs/hep-ph/0407333}{{\ttfamily hep-ph/0407333}}].

\bibitem{Huber:2007ji}
P.~Huber, J.~Kopp, M.~Lindner, M.~Rolinec and W.~Winter, \emph{{New features in
  the simulation of neutrino oscillation experiments with GLoBES 3.0: General
  Long Baseline Experiment Simulator}},
  \href{https://doi.org/10.1016/j.cpc.2007.05.004}{\emph{Comput. Phys. Commun.}
  {\bfseries 177} (2007) 432}
  [\href{https://arxiv.org/abs/hep-ph/0701187}{{\ttfamily hep-ph/0701187}}].

\bibitem{Prob3pp}
``Prob3++: https://webhome.phy.duke.edu/~raw22/public/prob3++/.''

\bibitem{Wallraff:2014qka}
M.~Wallraff and C.~Wiebusch, \emph{{Calculation of oscillation probabilities of
  atmospheric neutrinos using nuCraft}},
  \href{https://doi.org/10.1016/j.cpc.2015.07.010}{\emph{Comput. Phys. Commun.}
  {\bfseries 197} (2015) 185}
  [\href{https://arxiv.org/abs/1409.1387}{{\ttfamily 1409.1387}}].

\bibitem{GAMBIT:2017yxo}
{\scshape GAMBIT} collaboration, \emph{{GAMBIT: The Global and Modular
  Beyond-the-Standard-Model Inference Tool}},
  \href{https://doi.org/10.1140/epjc/s10052-017-5321-8}{\emph{Eur. Phys. J. C}
  {\bfseries 77} (2017) 784}
  [\href{https://arxiv.org/abs/1705.07908}{{\ttfamily 1705.07908}}].

\bibitem{Scott:2012qh}
P.~Scott, \emph{{Pippi - painless parsing, post-processing and plotting of
  posterior and likelihood samples}},
  \href{https://doi.org/10.1140/epjp/i2012-12138-3}{\emph{Eur. Phys. J. Plus}
  {\bfseries 127} (2012) 138}
  [\href{https://arxiv.org/abs/1206.2245}{{\ttfamily 1206.2245}}].

\bibitem{Gazizov:2004va}
A.~Gazizov and M.P.~Kowalski, \emph{{ANIS: High energy neutrino generator for
  neutrino telescopes}},
  \href{https://doi.org/10.1016/j.cpc.2005.03.113}{\emph{Comput. Phys. Commun.}
  {\bfseries 172} (2005) 203}
  [\href{https://arxiv.org/abs/astro-ph/0406439}{{\ttfamily
  astro-ph/0406439}}].

\bibitem{Vincent:2017svp}
A.C.~Vincent, C.A.~Arg\"uelles and A.~Kheirandish, \emph{{High-energy neutrino
  attenuation in the Earth and its associated uncertainties}},
  \href{https://doi.org/10.1088/1475-7516/2017/11/012}{\emph{JCAP} {\bfseries
  11} (2017) 012} [\href{https://arxiv.org/abs/1706.09895}{{\ttfamily
  1706.09895}}].

\bibitem{Garcia:2020jwr}
A.~Garcia, R.~Gauld, A.~Heijboer and J.~Rojo, \emph{{Complete predictions for
  high-energy neutrino propagation in matter}},
  \href{https://doi.org/10.1088/1475-7516/2020/09/025}{\emph{JCAP} {\bfseries
  09} (2020) 025} [\href{https://arxiv.org/abs/2004.04756}{{\ttfamily
  2004.04756}}].

\bibitem{Safa:2021ghs}
I.~Safa, J.~Lazar, A.~Pizzuto, O.~Vasquez, C.A.~Arg\"uelles and
  J.~Vandenbroucke, \emph{{TauRunner: A public Python program to propagate
  neutral and charged leptons}},
  \href{https://doi.org/10.1016/j.cpc.2022.108422}{\emph{Comput. Phys. Commun.}
  {\bfseries 278} (2022) 108422}
  [\href{https://arxiv.org/abs/2110.14662}{{\ttfamily 2110.14662}}].

\bibitem{Garg:2022ugd}
D.~Garg et~al., \emph{{Neutrino propagation in the Earth and emerging charged
  leptons with nuPyProp}},
  \href{https://doi.org/10.1088/1475-7516/2023/01/041}{\emph{JCAP} {\bfseries
  01} (2023) 041} [\href{https://arxiv.org/abs/2209.15581}{{\ttfamily
  2209.15581}}].

\bibitem{anaconda}
\emph{Anaconda software distribution},  2020.

\bibitem{pypi}
``Python package index - pypi.''

\end{thebibliography}\endgroup

\end{document}